\documentclass{IEEEtran}
\usepackage{cite}
\usepackage{graphicx}
\usepackage{amsmath}
\usepackage{array}
\usepackage{mdwmath}
\usepackage{amssymb}
\usepackage{mdwtab}
\usepackage{stfloats}
\usepackage[tight,footnotesize]{subfigure}
\usepackage{amsmath,amsthm}
\usepackage{threeparttable}
\usepackage{color}
\usepackage{algpseudocode}
\usepackage{algorithm}
\algrenewcommand{\algorithmicrequire}{\textbf{Input:}}
\algrenewcommand{\algorithmicensure}{\textbf{Output:}}

\newtheorem{example}{Example}

\newtheorem{remark}{Remark}
\newtheorem{lemma}{Lemma}

\begin{document}

\title{Polarization Shift Keying (PolarSK): System Scheme and Performance Analysis}
\author{Jiliang~Zhang,~\textit{Member~IEEE},~Yang~Wang,~Jie Zhang,~\textit{Senior~Member~IEEE},\\and~Liqin~Ding,~\textit{Member~IEEE}\
\thanks{Jiliang Zhang is with the  School of Information Science and Engineering, Lanzhou University, Lanzhou 730000, P. R. China.
Yang Wang and Liqin Ding are with Shenzhen Graduate School, Harbin Institute of Technology, Shenzhen 518055, P. R. China. Jie Zhang is with the Department of Electronic and Electrical Engineering, The University of Sheffield, Sheffield S1 3JD, UK. Corresponding Author: Yang Wang (yangw@hit.edu.cn)}
\thanks{The research is funded in part by National Natural Science Foundation of China (61501137,61371101), in part by the European Union's Horizon 2020 research and innovation programme-is3DMIMO (734798), and in part by the Fundamental Research Funds for the Central Universities (lzujbky-2017-38).
}
\thanks{Copyright (c) 2015 IEEE. Personal use of this material is permitted. However, permission to use this material for any other purposes must be obtained from the IEEE by sending a request to pubs-permissions@ieee.org.}
}
\markboth{IEEE Transactions on Vehicular Technology, vol. 00, no. 00, 2017}%
{Shell \MakeLowercase{\textit{et al.}}: Bare Demo of IEEEtran.cls for Journals}
\maketitle

\begin{abstract}
 The single-radio-frequency (RF) multiple-input-multiple-output (MIMO)  system has been proposed to pursue a high spectral efficiency
while keeping a low hardware cost and complexity.
Recently, the available degrees of freedom (DoF) in the polarization domain has been exploited to reduce the size of the transmit antenna array and provide 1 bit per channel use (bpcu) multiplexing gain for the single-RF MIMO  system. 
Nevertheless, the polarization domain resource still has the potential to provide a higher multiplexing gain in the polarized single-RF MIMO system. In this paper,
we propose a generalized polarization shift keying (PolarSK) modulation scheme that uses polarization states in the dual-polarized transmit antenna as an information-bearing unit to increase the overall spectral efficiency. At the receiver, the optimum maximum likelihood (ML) detector is employed to investigate the ultimate performance limit of the PolarSK. A closed-form union upper bound on the average bit error probability (ABEP) of the PolarSK with the optimum ML receiver is derived. 
Inspired by the analytic ABEP, a constellation diagram optimization algorithm is proposed.
To reduce the computational complexity of the receiver, a linear successive interference cancellation (SIC) detection algorithm and a sphere-decoding (SD) detection algorithm are introduced. Through numerical results, performances of the proposed PolarSK in terms of ABEP and computational complexity are analyzed. Furthermore, the PolarSK is analyzed over measured indoor channels. Numerical and measurement results show that the PolarSK scheme outperforms the state of the art dual-polarized/uni-polarized SM schemes.
\end{abstract}


\IEEEpeerreviewmaketitle

\section{Introduction}

Research works on future wireless communication technologies will be pursuing high data rate and spectral efficiency over the next decade  \cite{Cisco2}. Multi-antennas have been universally employed to enhance reliability and/or spectrum efficiency of wireless communication systems \cite{s1}. However, when multiple antenna elements work together, multiple expensive transmit radio frequency (RF) chains have to be equipped at the transmitter. As a state-of-the-art multi-antennas system, single-RF Multiple-Input-Multiple-Output (MIMO) schemes, where only one RF chain is used at the transmitter, have been proposed to reduce the complexity and cost of multiple-antenna systems  {\cite{survey,survey1,survey2,survey3}}.

Currently, the concept of the single-RF MIMO scheme covers the spatial modulation (SM) \cite{s2,s4,s5,s6,s10}, the space shift keying (SSK) \cite{s3,s11}, and the electronically steerable parasitic antenna radiators (ESPAR) \cite{espar1,espar2}  schemes. The single-RF MIMO scheme first appeared in the form of SM.
The main idea of the SM is to match a block of information bits onto both a constellation point in the signal domain and a transmit antenna index in the spatial
domain.
In each time slot, only one transmit antenna is activated by an RF switch. 
Theoretical analysis and simulation results show that the SM scheme outperforms conventional MIMO spatial multiplexing (SMX) schemes  {\cite{s9}}. Under certain configurations, the SM even has the potential to outperform MIMO spatial diversity schemes in terms of reliability {\cite{s2}}. It has also been proved that the SM is more robust than traditional MIMO systems in the presence of channel estimation error \cite{channele}, and the keyhole effect \cite{keyholeSM}.
In the SSK scheme, the antenna index is the only means to convey information so that the detection complexity of the SSK is lower than the SM. Also, the simplicity of the SSK provides the ease of integration.
As another type of single-RF MIMO scheme, the ESPAR emerges to improve the energy efficiency by controlling the effective coupling matrix. In the ESPAR, a single radiating element is surrounded by a number of parasitic elements. Reactances of the ESPAR array are controlled electronically to convey information in the beam domain.
 Some improved single-RF MIMO systems have also been reported \cite{im1,im2,im3,s13,s14,im5,im6,im7,signalconsSM, ES1,ES2,TWRCSM,shuaishuai,shuaishuai2,shuaishuai3}, and  the performance of single-RF MIMO systems was evaluated through analytical analysis {\cite{MDR1,MDR2,MDR3,bepsm,MDRSSK1,MDRSSK2,MDRSSK3,s7,MDRSSK4,SManalysis,mlsm}} and measurement  {\cite{mea1,mea2,mea3,mea4}}. Antenna arrays using semiconductor diodes which were proposed in \cite{switch} with a switching time of less than 100ns further facilitate the implementation of the single-RF MIMO scheme.

In multiple antenna systems, the physical space limitation to place multiple transmit and receive antennas is not always met by uni-polarized (UP) antennas, which require a large inter-antenna separation to reduce the correlation effect \cite{DPM2017}.
The available degrees of freedom (DoF) in the polarization domain has been exploited in conventional MIMO systems to reduce the size of the transmit antenna array {\cite{ploarc9,ploarc11,ploarc12,TITO}}. In practice, cellular land mobile radio systems have employed dual-polarized 45$^{\circ}$ slant antennas at base stations.
For two-input two-output (TITO)/two-input multiple-output (TIMO) systems, \cite{TITO} added the polarization dimension to the transceiver architecture to benefit from polarization diversity. However, the polarization domain DoF has rarely been exploited in single-RF system in the current literature.

In \cite{ICCPSM,WPCPSM}, a dual-polarized (DP)-SM scheme was proposed, where a selection of DP transmit antenna is employed in the SM to carry information. In \cite{DPM2017}, the bit error performance of the DP-SM was analyzed over correlated fading channels. However, even though the idea in \cite{ICCPSM,WPCPSM} is promising, the achievable spectral efficiency is limited by only one bit per channel use (bpcu) since only two orthogonal polarization states
are used.

The polarization domain DoF has the potential to provide a higher multiplexing gain since a DP transmit antenna can generate various polarization states including linear polarizations, circular polarizations and elliptic polarizations. Therefore, in this paper, we generalize the existing DP-SM scheme into a novel modulation scheme, referred to as polarization shift keying (PolarSK) by fully exploiting the polarization domain DoF to achieve a higher spectral efficiency. Main contributions of this paper are summarized as follows.

\begin{figure}[!t]
  \centering
  \includegraphics [width=3.5in]{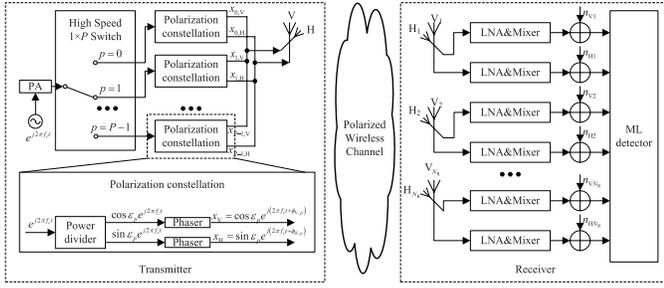}
  \centering\caption{The scheme of the PolarSK system.} \label{systemf}
  \vspace{-10pt}
\end{figure}

\begin{enumerate}
    \item As shown in Fig. 1, a novel polarized single-RF MIMO scheme, referred to as the PolarSK, is proposed considering a general set of polarization states. In the PolarSK, blocks of information bits are mapped onto the indices of different polarization states, e. g., linear polarizations, circular polarizations and elliptic polarizations, in one DP transmit antenna. Compared with the DP-SM shown in \cite{ICCPSM,WPCPSM,DPM2017}, a higher spectral efficiency is achieved.
    \item The optimal maximum likelihood (ML) PolarSK receiver is introduced to investigate the ultimate performance limit. An upper bound on the average bit error probability (ABEP) of the PolarSK over Rayleigh fading channels is derived in a closed-form taking advantage of Gray code bit mapping features.
    \item On the basis of the analytic ABEP upper bound, a constellation optimization scheme is designed by minimizing the maximum pairwise error probability (PEP). The optimality is verified by typical examples.
    \item In order to reduce the computational complexity of the PolarSK receiver, computational detection algorithms, i.e., the linear successive interference cancellation (SIC) and the sphere-decoding (SD) algorithms, are proposed to provide a tradeoff between computational complexity and ABEP performance. 
    \item The performance of the proposed PolarSK is evaluated  through numerical simulations and practical measurements. Both the computational complexity and the ABEP are taken into account. Numerical results show that the proposed PolarSK scheme outperforms state-of-the-art DP-SM and UP TITO/TIMO-SM schemes.
\end{enumerate}

The rest of this paper is organized as follows. 
Section II provides the transmission scheme of PolarSK. In Section III, analytic expressions of the ABEP of PolarSK are derived in closed-forms. On the basis of analytical results, in Section IV, we optimize the signal constellation in terms of ABEP. In Section V, computational detection algorithms of the PolarSK system are proposed. Finally, Section VI provides our numerical results, and Section VII concludes this paper.

\section{Transmission scheme}

Compared with the DP-SM scheme that uses vertical and horizontal polarizations only, the Stokes space {\cite{optsto}} is more efficient for high data rate transmission. Driven by this idea, the PolarSK system, considering a generic $1\times N_{\mathrm{R}}$ SIMO system with one DP transmit antenna and $N_{\mathrm{R}}$ DP receive antennas, is proposed in Fig. \ref{systemf}.

At the transmitter, the PolarSK system employs a size-$P$ polarization constellation $\{\mathbf{x}_0, \mathbf{x}_1,\cdots, \mathbf{x}_{P-1}\}$ with an average symbol power of $\mathrm{E}[|\mathbf{x}_p|^2]={1}$. The transmitter encodes a block of $\log_2 P$ data bits onto the indices of polarization states (e. g. linear polarizations, circular polarizations and elliptic polarizations), which are selected to transmit the carrier wave signal. The data rate of the PolarSK is $R=\log_2P$ bpcu.
In this paper, polarization states are represented in the form of the Jones vector in a right-handed Cartesian coordinate system \cite[eq. (1)]{bincao}:
\begin{eqnarray}
\label{signaldefine}
\mathbf{x}_p=
\left[
\begin{array}{l}
x_{p,\mathrm{V}}
\\
x_{p,\mathrm{H}}
\end{array}
\right]=
\left[
\begin{array}{l}
\cos\epsilon_{p}\exp\left(j\phi_{p,\mathrm{V}}\right)\\
\sin\epsilon_{p}\exp\left(j\phi_{p,\mathrm{H}}\right)
\end{array}
\right],
\end{eqnarray}
where $p=0,1,2,...,P-1$ is the index of polarization state, $\epsilon_{p}$, $\phi_{p,\mathrm{V}}$ and $\phi_{p,\mathrm{H}}$ are polarization parameters which determine the polarized states of the signal, $\epsilon_{p}$ denotes the polarized angle, and $\phi_{p,\mathrm{V}}$ and $\phi_{p,\mathrm{H}}$ denote the phase of the signal transmitted by the vertically and horizontally polarized antenna, respectively.
$\phi_{p,\mathrm{V}}-\phi_{p,\mathrm{H}}=0,\pi$ or $\epsilon_{p}=0,\pi/2$ denotes the linear polarization.
$\phi_{p,\mathrm{V}}-\phi_{p,\mathrm{H}}=\pi/2$ and $\epsilon_{p}=\pi/4$ indicates the left-handed circular polarization.
$\phi_{p,\mathrm{V}}-\phi_{p,\mathrm{H}}=-\pi/2$ and $\epsilon_{p}=\pi/4$ shows the right-handed circular polarization.

Particularly, in a DP-SM system with an $M$-phase shift keying (PSK) signal constellation \cite{ICCPSM,WPCPSM,DPM2017}, the candidate polarization state $\mathbf{x}_p$ is restricted to be in the subset
$\left\{
\left(\begin{smallmatrix}
\exp(\frac{jq}{2\pi})
\\
0
\end{smallmatrix}\right)
,
\left(\begin{smallmatrix}
0
\\
\exp(\frac{jq}{2\pi})
\end{smallmatrix}\right)
|q=1,2,...,M\right\}$. Thus, the DP-SM system can be seen as a special case of the proposed PolarSK system.

\begin{figure} [!t]
\centering
\subfigure[C$_1$,
R=6bpcu]{\includegraphics[width=2.7in]{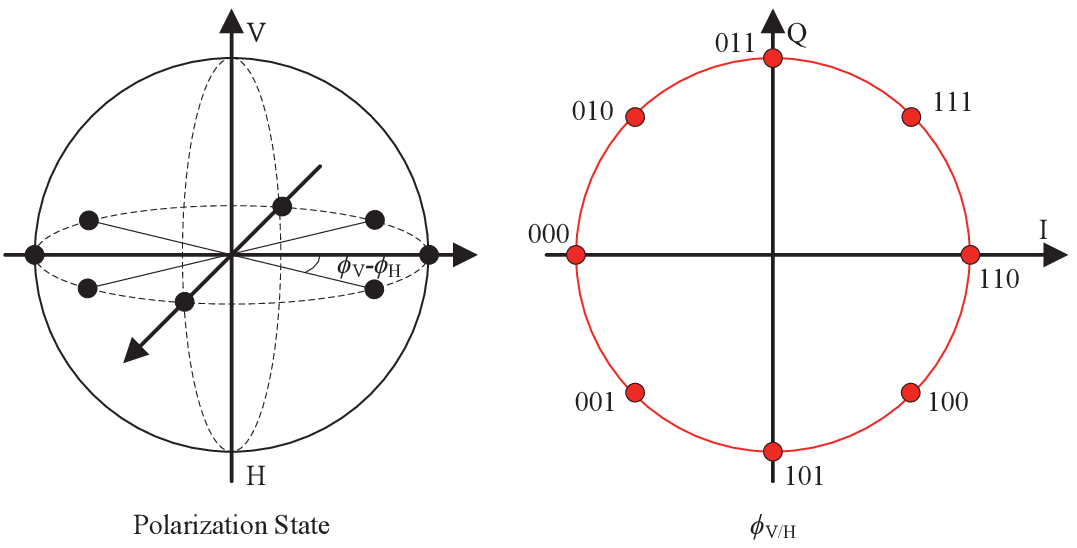}}
\subfigure[C$_2$,
R=7bpcu]{\includegraphics[width=2.7in]{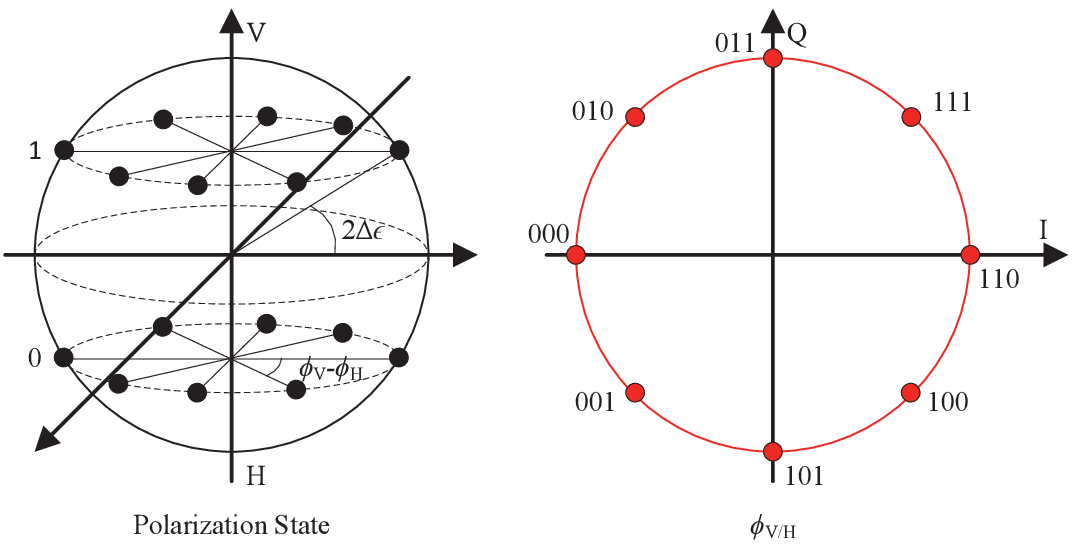}}
\caption{Signal points constellations for some examples of PolarSK modulation schemes.
}\label{examples}
 \vspace{-10pt}
\end{figure}

\begin{example}
To constrain the focus of this paper on proposing the PolarSK scheme, we use examples of signal constellations of the PolarSK shown in (\ref{stoks}).
\begin{eqnarray}
\label{stoks}
\mathbf{x}_{k,q_{\mathrm{V}},q_{\mathrm{H}}}=
\left[
\begin{array}{l}
\cos\epsilon_{k}\exp\left(j\frac{2\pi q_{\mathrm{V}}}{ M}\right)\\
\sin\epsilon_{k}\exp\left(j\frac{2\pi q_{\mathrm{H}}}{ M}\right)
\end{array}
\right],
\end{eqnarray}
where $q_{\mathrm{V}},q_{\mathrm{H}}=1,2,...,M$ are phase indices of signals that are respectively transmitted by vertical and horizontal transmit antennas. $q_{\mathrm{V}}$ and $q_{\mathrm{H}}$ are both Gray mapped following $M$-PSK rule. $k=1,2,...,K$ is the index of $\epsilon_{p}$. Constellation points are distributed in $K$ circles of latitude with identical interval in the  the Poincar$\acute{e}$ sphere. In the rest of this paper, the constellation diagram is denoted as $C_K$, and the data rate of the $C_K$ constellation is $R=\log_2(M^2K)$.
\end{example}

\begin{remark}
For a constellation diagram that is defined in Example 1, the data rate of the signal points constellation C$_K$ is $R=\log_2P=\log_2M^2K$bpcu. Taking bandwidth into account, the data rate of the transmission using a root raised cosine (RRC) filter is \cite[Eq. (7)]{pulse}
\begin{eqnarray}
R_{\mathrm{s}}=\frac{T_{\mathrm{s}}\log_2M^2K}{N_{\mathrm{s}}(1+\alpha)}
\mathrm{[bits/s/Hz]},
\end{eqnarray}
where $N_{\mathrm{s}}$ denotes the number of samples in one pulse, and $T_{\mathrm{s}}$ is the symbol period constrained by the switching time and the Nyquist criterion. If the improved $D$ RF chains aided approach proposed in \cite{pulse} is employed, the data rate will be $D\times R_{\mathrm{s}}$.
\end{remark}

\begin{figure*}[!t]
\begin{eqnarray}
\setcounter{equation}{11}
\label{peprayeq5}
&&\hspace{-0.5in}
\begin{array}{l}
\mathrm{Pr}(p\rightarrow \hat{p})
=
\frac{
\left[\left(1+ \frac{\rho\Lambda_{\mathrm{V}}}{4}\right)\left(1+ \frac{\rho\Lambda_{\mathrm{H}}}{4}\right)\right]^{-N_{\mathrm{R}}}
\mathcal{B}\left(\frac{1}{2},2N_{\mathrm{R}}+\frac{1}{2}\right)
{}F_1\left[\frac{1}{2},N_{\mathrm{R}},N_{\mathrm{R}},2N_{\mathrm{R}}+1,(1+\frac{\rho\Lambda_{\mathrm{V}}}{4})^{-1},(1+\frac{\rho\Lambda_{\mathrm{H}}}{4})^{-1}\right]
}{2\pi}.
\end{array}
\\
\label{appell}
&&\hspace{-0.5in}
\begin{array}{l}
F_1(a,b_1,b_2,c;z_1,z_2)\triangleq \sum\limits_{d_1=0}^\infty\sum\limits_{d_2=0}^\infty \frac{(a)_{d_1+d_2} (b_1)_{d_1} (b_2)_{d_2}z_1^{d_1} z_2^{d_2}} {(c)_{d_1+d_2} d_1!d_2!}
\equiv
\frac{\Gamma(c)} {\Gamma(a)\Gamma(c-a)}
\int_0^1 t^{a-1} (1-t)^{c-a-1} (1-z_1t)^{-b_1} (1-z_2t)^{-b_2} \mathrm{d}t.
\end{array}
\\
\setcounter{equation}{22}
\label{peprayeq6}
&&\hspace{-0.4in}
\mathrm{Pr}(p\rightarrow \hat{p})
\doteq
\frac{\left[\left(1+ \frac{\rho\Lambda_{\mathrm{V}}}{4}\right)\left(1+ \frac{\rho\Lambda_{\mathrm{H}}}{4}\right)\right]^{-N_{\mathrm{R}}}
\mathcal{B}\left(\frac{1}{2},2N_{\mathrm{R}}+\frac{1}{2}\right)}{2\pi}
\doteq\frac{\left(\frac{\Lambda_{\mathrm{V}}\Lambda_{\mathrm{H}}}{16}\right)^{-N_{\mathrm{R}}}
\mathcal{B}\left(\frac{1}{2},2N_{\mathrm{R}}+\frac{1}{2}\right)}{2\pi}\rho^{-2N_{\mathrm{R}}}.
\end{eqnarray}
\hrulefill
  \vspace{-10pt}
\end{figure*}

\begin{figure}[!t]
  \centering
  \includegraphics [width=1.8in]{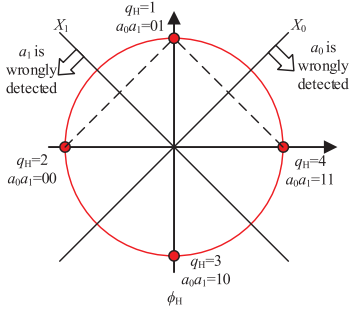}
  \centering\caption{{Signal-space diagram of $q_{\mathrm{H}}$ for $M=4$ and the error space when $q_{\mathrm{H}}=1$ is transmitted.} } \label{Meq4}
  \vspace{-10pt}
\end{figure}

The modulated signal vector is transmitted over one DP antenna and received by an array of $N_{\mathrm{R}}$ DP antennas. The channel can be modeled by a $2\times2N_{\mathrm{R}}$ dimensional channel matrix as

\begin{eqnarray}
\setcounter{equation}{4}
\label{channelmodeleq}
\begin{array}{l}
\mathbf{H}
=
\left[
\begin{array}{l l l l l l}
\mathbf{h}_{1}^{\mathrm{H}}&
\mathbf{h}_{2}^{\mathrm{H}}&
\cdots&
\mathbf{h}_{n_{\mathrm{R}}}^{\mathrm{H}}&
\cdots&
\mathbf{h}_{N_{\mathrm{R}}}^{\mathrm{H}}
\end{array}
\right]^{\mathrm{H}}
\\ \hspace{0.166in}=\left[
\begin{array}{l l}
\mathbf{h}_{\mathrm{V}}&\mathbf{h}_{\mathrm{H}}
\end{array}
\right]
,\end{array}
\end{eqnarray}
where $\circ^{\mathrm{H}}$ denotes Hermitian matrix transposition, and
\begin{eqnarray}
\label{channelmodeleq2}
\mathbf{h}_{n_{\mathrm{R}}}
=
\left[
\begin{smallmatrix}
  h_{\mathrm{VV},n_{\mathrm{R}}}
& h_{\mathrm{VH},n_{\mathrm{R}}}\sqrt{X}\\
  h_{\mathrm{HV},n_{\mathrm{R}}}\sqrt{X}
& h_{\mathrm{HH},n_{\mathrm{R}}}
\end{smallmatrix}
\right].
\end{eqnarray}
$X$ is combined by the imperfect cross-polar isolation (XPI) in the antenna array and the cross-polar ratio (XPR) in the propagation channel {\cite{TITO}}. In order to minimize the complexity of system modelling without loss of accuracy, we assign a constant value to $X$ \cite{channelmodel2,channelmodel3,channelmodel4}. We consider the frequency-flat slowly-varying fading channel model, where $h_{\mathrm{VV},n_{\mathrm{R}}}$, $h_{\mathrm{VH},n_{\mathrm{R}}}$, $h_{\mathrm{HV},n_{\mathrm{R}}}$, and $h_{\mathrm{HH},n_{\mathrm{R}}}$ are assumed to have uniformly distributed phases and {Rayleigh} distributed amplitudes.

The received signal of the PolarSK is expressed as
\begin{eqnarray}
\label{receivedsignal}
\mathbf{y}=\mathbf{H}\mathbf{x}_p+\frac{\mathbf{w}}{\sqrt{\rho}},
\end{eqnarray}
where $\rho\triangleq \frac{E_{\mathrm{s}}}{N_0}$ is the average signal to noise ratio (SNR), and $\mathbf{w}$ is the noise vector at the receiver following the independent and identically distributed (i.i.d.) complex Gaussian distribution {with a variance of 1 for each element}.

Under a specific $\mathbf{H}\mathbf{x}_p$, $\mathbf{y}$ is independent complex Gaussian distributed with the expectation $\mathbf{H}\mathbf{x}_p$. The optimum ML PolarSK receiver is the minimum Euclidean distance (MED) receiver \cite{mlsm,bepsm}, where the received message is the index of the $\mathbf{x}_{\hat{p}}$ that can minimize the Euclidean distance between $\rho\mathbf{Hx}_{\hat{p}}$ and the received signal, i.e.,
\begin{eqnarray}
\label{mlreceiver}
\hat{p}=\underset{{p}\in\{1,2,...,P\}}{\arg\min}\|\mathbf{y}-\mathbf{Hx}_{{p}}\|^2,
\end{eqnarray}
where $\|\circ\|$ denotes the 2-norm of a vector.

For a C$_K$ signal constellation proposed in Example 1, the ML decision rule in (\ref{mlreceiver}) is rewritten as
\begin{eqnarray}
\label{mlreceiver1}
[\hat{k},\hat{q}_{\mathrm{V}},\hat{q}_{\mathrm{H}}]=
\underset
{\begin{smallmatrix}{k}\in\{1,...,K\},\\
{q}_{\mathrm{V}}\in\{1,...,M\},\\
{q}_{\mathrm{H}}\in\{1,...,M\}
\end{smallmatrix}}
{\arg\min}\|\mathbf{y}-\mathbf{Hx}_{{k},{q}_{\mathrm{V}},{q}_{\mathrm{H}}}\|^2.
\end{eqnarray}

\section{ABEP of optimum receiver}
In this subsection, an analytic closed-form  ABEP upper bound of PolarSK systems are given over fading channels. Since the exact ABEP of (\ref{mlreceiver}) is infeasible to be derived in a closed-form, we employ the union upper bound technique, which is widely used to calculate the ABEP upper bound of single-RF MIMO systems \cite[Eq. (3)]{bepsm},\cite[Eq. (5)]{mlsm}.
Since the ABEP is irrelevant to the specific $q_{\mathrm{V}}$ or $q_{\mathrm{H}}$ that is transmitted due to the symmetry of the signal constellation, a union upper bound on the ABEP of the PolarSK system is computed by

\begin{eqnarray}
\label{ABEPeqst1}
\begin{array}{l}
ABEP
\leq\sum\limits_{k=1}^{K}\sum\limits_{\hat{k}=1}^{K} \sum\limits_{\hat{q}_{\mathrm{V}}=1}^{M}
ABEP_{k,\hat{k},\hat{q}_{\mathrm{V}}}
,
\end{array}
\end{eqnarray}
where $ABEP_{k,\hat{k},\hat{q}_{\mathrm{V}}}$ denotes the ABEP of the PolarSK system under the condition that ${q}_{\mathrm{V}}\rightarrow\hat{q}_{\mathrm{V}}$, $k\rightarrow\hat{k}$, ${q}_{\mathrm{H}}=1$ and ${q}_{\mathrm{V}}=1$, and following the union upper bound technique, we have
\begin{eqnarray}
\label{ABEPeqst2}
\begin{array}{l}
ABEP_{k,\hat{k},\hat{q}_{\mathrm{V}}}
\leq\frac{1}{K\log_2(KM^2)}\sum\limits_{\hat{q}_{\mathrm{H}}=1}^{M}
\\
\hspace{0.76in}\times
N\left[(k,1,1)\rightarrow (\hat{k},\hat{q}_{\mathrm{V}},\hat{q}_{\mathrm{H}})\right]
\\
\hspace{0.76in}\times \mathrm{Pr}\left[(k,1,1)\rightarrow (\hat{k},\hat{q}_{\mathrm{V}},\hat{q}_{\mathrm{H}})\right]
,
\end{array}
\end{eqnarray}
where the pairwise error probability (PEP)  $\mathrm{Pr}[(k,q_{\mathrm{V}},q_{\mathrm{H}})\rightarrow (\hat{k},\hat{q}_{\mathrm{V}},\hat{q}_{\mathrm{H}})]$ is defined as the probability of choosing the wrong symbol $(\hat{k},\hat{q}_{\mathrm{V}},\hat{q}_{\mathrm{H}})$ at the receiver assuming that there are only two symbols $(k,q_{\mathrm{V}},q_{\mathrm{H}})$ and $(\hat{k},\hat{q}_{\mathrm{V}},\hat{q}_{\mathrm{H}})$ possibly being transmitted, and the $N[(k,q_{\mathrm{V}},q_{\mathrm{V}})\rightarrow (\hat{k},\hat{q}_{\mathrm{V}},\hat{q}_{\mathrm{H}})]$ denotes the Hamming distance between $(k,q_{\mathrm{V}},q_{\mathrm{H}})$ and $(\hat{k},\hat{q}_{\mathrm{V}},\hat{q}_{\mathrm{H}})$. 
The computation of the PEP will be provided in Lemma \ref{peplemma}.

\begin{lemma}
\label{peplemma}
Over i.i.d. Rayleigh fading channels, the analytic PEP is computed by (\ref{peprayeq5}), where $\mathcal{B}(a,b)\triangleq \frac{\Gamma(a)\Gamma(b)}{\Gamma(a+b)}$ denotes the Beta function \cite{betafunction}, $\Gamma(a)$ denotes the Gamma function, $F_1$ denotes the Appell hypergeometric function \cite{appellcite} defined as (\ref{appell}),
\begin{eqnarray}
\setcounter{equation}{13}
&&(a)_d \triangleq \frac{\Gamma(a+d)}{\Gamma(a)},
\\
\label{lambda1}
&&\Lambda_{\mathrm{V}}\triangleq |\Delta{x}_{\mathrm{V}}|^2+X|\Delta{x}_{\mathrm{H}}|^2,
\\
\label{lambda2}
&&\Lambda_{\mathrm{H}}\triangleq X|\Delta{x}_{\mathrm{V}}|^2+|\Delta{x}_{\mathrm{H}}|^2,
\\
&&
\left[
\begin{smallmatrix}\Delta{x}_{\mathrm{V}} \\ \Delta{x}_{\mathrm{H}}\end{smallmatrix}
\right]
\triangleq
\left[
\begin{smallmatrix}
\cos\epsilon_{k}\exp\left(j\frac{2\pi q_{\mathrm{V}}}{ M}\right)-\cos\epsilon_{\hat{k}}\exp\left(j\frac{2\pi \hat{q}_{\mathrm{V}}}{ M}\right)
\\
\sin\epsilon_{k}\exp\left(j\frac{2\pi q_{\mathrm{H}}}{ M}\right)-\sin\epsilon_{\hat{k}}\exp\left(j\frac{2\pi \hat{q}_{\mathrm{H}}}{ M}\right)
\end{smallmatrix}
\right].
\end{eqnarray}
\end{lemma}
\begin{IEEEproof} See Appendix \ref{prooflemma0appen}. \end{IEEEproof}

Taking advantage of Gray code bit mapping features of $q_{\mathrm{H}}$ and $q_{\mathrm{V}}$, the upper bound in Eq. (\ref{ABEPeqst2}) is tightened as follows.

Firstly, we begin by the simple case of $M=4$. The bit-mapping Gray code of $q_{\mathrm{H}}$ is denoted as $a_0a_1$, as in Fig. \ref{Meq4}. On the basis of the union upper bound approach, we obtain
\begin{eqnarray}
\label{ABEPeqst4}
\begin{array}{l}
ABEP_{k,\hat{k},\hat{q}_{\mathrm{V}}}
\leq\frac{N\left[(k,1,1)\rightarrow (\hat{k},\hat{q}_{\mathrm{V}},1)\right]
+\sum\limits_{i=0}^{1}P_{e,i}}{K\log_2(KM^2)},
\end{array}
\end{eqnarray}
where $P_{e,i}$ denotes the probably that the bit $a_i$ is wrongly detected. Clearly, it is observed in Fig. \ref{Meq4} that
\begin{eqnarray}
\label{ABEPeqst5}
\left\{
\begin{array}{l}
P_{e,0}\equiv \mathrm{Pr}\left[(k,1,1)\rightarrow (\hat{k},\hat{q}_{\mathrm{V}},4)\right],
\\
P_{e,0}\equiv \mathrm{Pr}\left[(k,1,1)\rightarrow (\hat{k},\hat{q}_{\mathrm{V}},2)\right].
\end{array}
\right.
\end{eqnarray}
Substituting (\ref{ABEPeqst5}) into (\ref{ABEPeqst4}), we have
\begin{eqnarray}
\label{ABEPeqst6}
\begin{array}{l}
ABEP_{k,\hat{k},\hat{q}_{\mathrm{V}}}
\leq\frac{1}{K\log_2(KM^2)}\sum\limits_{\hat{q}_{\mathrm{H}}=\{1,2,4\}}
\\
\hspace{0.76in}\times
N\left[(k,1,1)\rightarrow (\hat{k},\hat{q}_{\mathrm{V}},\hat{q}_{\mathrm{H}})\right]
\\
\hspace{0.76in}\times \mathrm{Pr}\left[(k,1,1)\rightarrow (\hat{k},\hat{q}_{\mathrm{V}},\hat{q}_{\mathrm{H}})\right].
\end{array}
\end{eqnarray}

Comparing (\ref{ABEPeqst2}) and (\ref{ABEPeqst6}), it is clearly observed (\ref{ABEPeqst6}) is a tighter upper bound of $ABEP_{k,\hat{k},\hat{q}_{\mathrm{V}}}$ than (\ref{ABEPeqst2}) since the item for $\hat{q}_{\mathrm{H}}=3$ is not summed. And therefore, $\hat{q}_{\mathrm{H}}=3$ is a \textit{redundant item} for ABEP calculation, which loosens the union upper bound.
Similarly, since the $q_{\mathrm{H}}$ and $q_{\mathrm{V}}$ are interchangeable while calculating the ABEP, $\hat{q}_{\mathrm{V}}=3$ is also a redundant item. Thus, for $M=4$, the upper bound is tightened as
\begin{eqnarray}
\label{ABEPeqst7}
\begin{array}{l}
ABEP\leq\frac{1}{K\log_2(KM^2)}\sum\limits_{k=1}^{K}\sum\limits_{\hat{k}=1}^{K} \sum\limits_{\hat{q}_{\mathrm{V}}=\{1,2,4\}} \sum\limits_{\hat{q}_{\mathrm{H}}=\{1,2,4\}}
\\
\hspace{0.5in}\times
N\left[(k,1,1)\rightarrow (\hat{k},\hat{q}_{\mathrm{V}},\hat{q}_{\mathrm{H}})\right]
\\
\hspace{0.5in}\times \mathrm{Pr}\left[(k,1,1)\rightarrow (\hat{k},\hat{q}_{\mathrm{V}},\hat{q}_{\mathrm{H}})\right]
.
\end{array}
\end{eqnarray}

Then, we generalize Eq. (\ref{ABEPeqst7}) for Gray coded PolarSK systems with $M>4$. The pathway to tighten the upper bound is to find out all redundant items, and remove them in summations.
\cite[Section III]{PSK} provided a pathway to find out redundant items of union upper bound, although its initial objective was to provide an approximate BEP of traditional $M$-PSK transmissions over AWGN channels. By using the similar approach in \cite[Section III]{PSK} and following the idea of selecting redundant items for $M=4$, it is found that the redundant items in PolarSK systems are $\hat{q}_{\mathrm{V}/\mathrm{H}}=\{3,5,7,...,M-1\}$. Therefore, the union upper bound on the ABEP is tightened as
\begin{eqnarray}
\label{ABEPeqst8}
\begin{array}{l}
ABEP\leq\frac{1}{K\log_2(KM^2)}
\sum\limits_{\hat{q}_{\mathrm{V}}=\{1,2:2:M\}} \sum\limits_{\hat{q}_{\mathrm{H}}=\{1,2:2:M\}}
\\
\hspace{0.5in}\times \sum\limits_{k=1}^{K}\sum\limits_{\hat{k}=1}^{K}
N\left[(k,1,1)\rightarrow (\hat{k},\hat{q}_{\mathrm{V}},\hat{q}_{\mathrm{H}})\right]
\\
\hspace{0.5in}\times \mathrm{Pr}\left[(k,1,1)\rightarrow (\hat{k},\hat{q}_{\mathrm{V}},\hat{q}_{\mathrm{H}})\right]
,
\end{array}
\end{eqnarray}
where $2:2:M\triangleq 2,4,6,...,M$.

\begin{algorithm}[!t]
\caption{Algorithm for computing the optimal $\Delta \epsilon_{\mathrm{opt}}$}
\label{NRalgo}
{\begin{algorithmic}[1]
\Require $K$, $M$, $X$, $N$
\Ensure $\Delta \epsilon_{\mathrm{opt}}$
\State $x \gets 0$, $n \gets 0$
\State $\Theta\gets\sqrt[4]{\frac{X\left(1-\cos\frac{2\pi}{M}\right)^2}{4(1+X)^2}}$
\For {$n=1$ to $N$}
    \State $f \gets \frac{\Theta\{\cos[(K-1)x]+\sin[(K-1)x]\}-\sin x}{\Theta\{(K-1)\sin[(K-1)x]-(K-1)\cos[(K-1)x]\}+\cos x}$
    \State $x \gets x+f$
\EndFor
\State $\Delta \epsilon_{\mathrm{opt}}\gets x$
\end{algorithmic}}
\end{algorithm}

To derive the diversity order of the proposed PolarSK, we define an operator $\doteq$ such that for a probability $f(\rho)$, if $\lim\limits_{\rho \rightarrow \infty} \frac{\log f(\rho)}{\log (\alpha\rho)} = -\mathcal{O}$ and $\mathcal{O}$ is a constant, then $\log f(\rho)\doteq (\alpha\rho)^{-\mathcal{O}}$.
Applying $F_1\left[\frac{1}{2},N_{\mathrm{R}},N_{\mathrm{R}},2N_{\mathrm{R}}+1,0,0\right]=1$, $\Lambda_{\mathrm{V}}> 0$ and $\Lambda_{\mathrm{H}}> 0$ in the high SNR regime, we obtain an asymptotic PEP as (\ref{peprayeq6}).

Substituting the asymptotic PEP (\ref{peprayeq6}) into the union upper bound (\ref{ABEPeqst8}), we obtain an asymptotic ABEP in the high SNR regime as
\begin{eqnarray}
\setcounter{equation}{23}
\label{ABEPfinal2}
\begin{array}{l}
ABEP\leq\frac{\mathcal{B}\left(\frac{1}{2},2N_{\mathrm{R}}+\frac{1}{2}\right) \mathcal{G}}{2\pi {16}^{-N_{\mathrm{R}}}\log_2(KM^2)\rho^{2N_{\mathrm{R}}}},
\end{array}
\end{eqnarray}
where
\begin{eqnarray}
\label{ABEPfinalG}
\begin{array}{l}
\mathcal{G}\triangleq
\frac{1}{K}
\sum\limits_{\hat{q}_{\mathrm{V}}=\{1,2:2:M\}} \sum\limits_{\hat{q}_{\mathrm{H}}=\{1,2:2:M\}}\sum\limits_{k=1}^{K}\sum\limits_{\hat{k}=1}^{K}
\\
\hspace{0.3in}\times 
N\left[(k,1,1)\rightarrow (\hat{k},\hat{q}_{\mathrm{V}},\hat{q}_{\mathrm{H}})\right]\left({\Lambda_{\mathrm{V}}\Lambda_{\mathrm{H}}}\right)^{-N_{\mathrm{R}}}.
\end{array}
\end{eqnarray}

Eq. (\ref{ABEPfinal2}) provides a pathway to analyze the diversity order and coding gain of the PolarSK, as shown in Remark \ref{gainanalysis}.
\begin{remark}
\label{gainanalysis}
As shown in Eq. (\ref{ABEPfinal2}), because the corresponding ABEP is proportional to $\rho^{-2N_{\mathrm{R}}}$ in the high SNR regime, the diversity order of the PolarSK system is $2 N_{\mathrm{R}}$. Consequently, the coding gain is computed by 
$$\mathcal{C}=4\left(\frac{2\pi \log_2(KM^2)}{\mathcal{B}\left(\frac{1}{2},2N_{\mathrm{R}}+\frac{1}{2}\right) \mathcal{G}}\right)^{\frac{1}{2N_{\mathrm{R}}}},$$
and  $ABEP\doteq (\mathcal{C}\rho)^{-2N_{\mathrm{R}}}$.

Furthermore, Eqs. (\ref{ABEPfinal2}-\ref{ABEPfinalG}) indicate that given an $N_{\mathrm{R}}$ and an SNR, the asymptotic upper bound on the ABEP depends on the minimum $\Lambda_{\mathrm{V}}\Lambda_{\mathrm{H}}$ for all candidate polarization states. Therefore, we can optimize the signal constellation of the PolarSK by minimizing $\Lambda_{\mathrm{V}}\Lambda_{\mathrm{H}}$ to achieve the optimum ABEP performance. The optimization is given in the following section.
\end{remark}

\section{Optimization of C$_K$ constellation}
\label{optimizesection}

The constellation optimization problem in conventional SM systems was addressed in \cite{shuaishuai,shuaishuai2,shuaishuai3}. In this Section, the PolarSK constellation optimization problem is stated and addressed. 
For C{$_K$} signal constellation, the selection of candidate $\epsilon_k$ in Eq. (\ref{stoks}) determines the reliability of detecting both $k$ and $q$. We use two extreme situations to show the importance of optimizing $\vec{\epsilon}$.
If $\Delta \epsilon\triangleq\min\limits_{k_1,k_2=1,2,...,K, k_1\neq k_2}\{0.5|\epsilon_{k_1}-\epsilon_{k_2}|\}$ is too small, two plates of signal points will be very close to a same  circle of latitude of the Poincar$\rm{\acute{e}}$ sphere, so that a pair of signal points with a same phase but different $\epsilon_k$ will be difficult to detect and the estimation error of $\hat{k}$ will dominate the ABEP performance. Else if $\Delta \epsilon$ is too large, the plates of signal points will be respectively zoomed at V and H poles, and the estimation error of $\hat{q}$ will dominate the ABEP performance. Therefore, it is natural to ask what is the optimal $\vec{\epsilon}$ to maximize the ABEP. As the ABEP performance is dominated by the maximum PEP, the optimum {$\epsilon_k$} should be selected as
\begin{eqnarray}
\label{optcondition}
\begin{array}{l}
\epsilon_{k,\mathrm{opt}}=
\arg \min\limits_{\epsilon_k} \max\limits_{\hat{k},\hat{q}_{\mathrm{V}},\hat{q}_{\mathrm{H}}} \{PEP(k,q_{\mathrm{V}},q_{\mathrm{H}},\hat{k},\hat{q}_{\mathrm{V}},\hat{q}_{\mathrm{H}})\}\\
\hspace{0.36in}=\arg \max\limits_{\epsilon_k} \min\limits_{\hat{k},\hat{q}_{\mathrm{V}},\hat{q}_{\mathrm{H}}} \{\Lambda_{\mathrm{V}}\Lambda_{\mathrm{H}}\}.
\end{array}
\end{eqnarray}
The optimal $\epsilon_k$ is obtained by using Lemma \ref{d1}.
\begin{lemma}
\label{d1}
The optimal $\Delta \epsilon_{\mathrm{opt}}$ for PolarSK signal constellations C$_K$ is the minimum positive real root of Eq. (\ref{finaloptimal}), which can be obtained by Newton-Raphson method \cite{NRM}.
\begin{eqnarray}
\label{finaloptimal}
\Theta\{\cos[(K-1)x]+\sin[(K-1)x]\}=\sin x,
\end{eqnarray}
where
\begin{eqnarray}
\Theta\triangleq\sqrt[4]{\frac{X\left(1-\cos\frac{2\pi}{M}\right)^2}{4(1+X)^2}}.
\end{eqnarray}
Especially, when $K=2$, the root of Eq. (\ref{finaloptimal}) has a closed-form as
\begin{eqnarray}
\label{finaloptimalc2}
\begin{array}{l}
\Delta\epsilon_{\mathrm{opt}}
=\arctan\left(\frac{\Theta}{1+\Theta}\right).
\end{array}
\end{eqnarray}
{For $K>2$, since $\frac{\partial (\Theta\{\cos[(K-1)x]+\sin[(K-1)x]\}-\sin x)}{\partial x}=\Theta\{-(K-1)\sin[(K-1)x]+(K-1)\cos[(K-1)x]\}-\cos x$, $\Delta \epsilon_{\mathrm{opt}}$ is computed by Algorithm \ref{NRalgo}.}
\end{lemma}

\begin{IEEEproof}
See Appendix \ref{prooflemma1appen}.

\end{IEEEproof}


\section{Computational detection algorithms}

\begin{algorithm}[!t]
\caption{Pseudo-code for the QR-aided ML algorithm.}
\label{MLalgo}
\begin{algorithmic}[1]
{
\Require $\mathbf{H}$, $\mathbf{y}$
\Ensure $\hat{k}$,$\hat{q}_{\mathrm{V}}$,$\hat{q}_{\mathrm{H}}$
\State $[\mathbf{Q},\mathbf{R}] \gets \mathrm{qr}(\mathbf{H})$, $\mathbf{q}_1 \gets \mathbf{Q}(:,1)^{\mathrm{H}}$, $\mathbf{q}_2 \gets \mathbf{Q}(:,2)^{\mathrm{H}}$, $r_1 \gets \mathbf{R}(1,1)$, $r_{12} \gets \mathbf{R}(1,2)$, $r_2 \gets \mathbf{R}(2,2)$, $\tilde{y}_1 \gets \mathbf{q}_1\mathbf{y}$, $\tilde{y}_2 \gets \mathbf{q}_2\mathbf{y}$
\State $d_{\mathrm{s}} \gets +\infty$
\For {$k_0=1$ to $K$}
    \For {$q_{\mathrm{H},0}=1$ to $M$}
        \For {$q_{\mathrm{V},0}=1$ to $M$}
            \State $d_{\mathrm{s}}' \gets
               |\tilde{y}_1-r_1\cos\epsilon_{k_0}\exp(j\frac{2\pi q_{\mathrm{V},0}}{M})-r_{12}\sin\epsilon_{k_0}\exp(j\frac{2\pi q_{\mathrm{H},0}}{M})|^2+|\tilde{y}_2-r_{2}\sin\epsilon_{k_0}\exp(j\frac{2\pi q_{\mathrm{H},0}}{M})|^2$
            \If{$d_{\mathrm{s}}' <d_{\mathrm{s}}$}
            \State $\hat{q}_{\mathrm{H}} \gets q_{\mathrm{H},0}$
            \State $\hat{q}_{\mathrm{V}} \gets q_{\mathrm{V},0}$
            \State $\hat{k} \gets k_0$
            \State $d_{\mathrm{s}}\gets d_{\mathrm{s}}'$
            \EndIf
        \EndFor
   \EndFor
\EndFor
}
\end{algorithmic}
\end{algorithm}

It is observed from Eq. (\ref{mlreceiver1}) that when the number of receive antennas is large, the ML receiver is computationally expensive because $\mathbf{Hx}_{{k},{q}_{\mathrm{V}},{q}_{\mathrm{H}}}$ has to be computed once for each candidate symbol per channel use. In this section, we aim to propose simple receiver algorithms for the PolarSK.

\subsection{QR-aided ML detection}
Here, we start by introducing a QR-aided ML detection algorithm, whose computational complexity is not sensitive to $N_{\mathrm{R}}$, as a benchmark for complexity analysis. To further facilitate the PolarSK signal detection, two computational receivers will be proposed in the following context in this section.

Considering the QR factorization of the channel matrix  $\mathbf{H}$, we have
\begin{eqnarray}
\label{qrofh}
\mathbf{H}=\mathbf{QR}=[\mathbf{q}_1^{\mathrm{H}},\mathbf{q}_2^{\mathrm{H}},...,\mathbf{q}_{2N_{\mathrm{T}}}^{\mathrm{H}}]\left[\begin{smallmatrix}r_1&r_{12}\\0&r_2\\0&0\\...&...\\0&0\end{smallmatrix}\right].
\end{eqnarray}

Substituting Eq. (\ref{qrofh}) into Eq. (\ref{mlreceiver1}) and following some algebraic manipulations, the ML decision rule is rewritten as
\begin{eqnarray}
\label{mlck}
[\hat{k},\hat{q}_{\mathrm{V}},\hat{q}_{\mathrm{H}}]=
\underset
{{k},{q}_{\mathrm{V}},{q}_{\mathrm{H}}}{\arg\min}
\left\{
\begin{smallmatrix}
|\tilde{y}_1-r_1\cos\epsilon_{{k}}\exp(j\frac{2\pi {q}_{\mathrm{V}}}{ M})\\
-r_{12}\sin\epsilon_{{k}}\exp(j\frac{2\pi {q}_{\mathrm{H}}}{ M})|^2\\
+|\tilde{y}_2-r_{2}\sin\epsilon_{{k}}\exp(j\frac{2\pi {q}_{\mathrm{H}}}{ M})|^2
\end{smallmatrix}
\right\},
\end{eqnarray}
where $\tilde{y}_1\triangleq\mathbf{q}_1\mathbf{y}$ and $\tilde{y}_2\triangleq\mathbf{q}_2\mathbf{y}$. On the basis of (\ref{mlck}), the pseudo-code of the QR-aided ML detection algorithm is shown in Algorithm \ref{MLalgo}.

For the QR-ML detection algorithm, the detection of $\hat{k}$  leads to a loop given in lines (3-13) of Algorithm 2. Compared to the conventional QR-ML algorithm in 
the UP-TIMO scheme, the loop increases the computational complexity. However, it is observed that the computational complexity of the loops is not sensitive to $N_{\mathrm{R}}$.
The only part of Algorithm \ref{MLalgo} whose computational complexity is related to $N_{\mathrm{R}}$ is the QR factorization of the channel matrix.
Over fast fading channels, the operation of big matrix, i.e. the QR factorization, is required to be conducted only once per channel use, whereas for conventional ML receiver, the operation of big matrix, i.e., $\mathbf{Hx}_{{k},{q}_{\mathrm{V}},{q}_{\mathrm{H}}}$, has to be conducted $KM^2$ times.
Therefore, it is predictable that the QR-aided ML detection algorithm has a lower computational complexity than the conventional ML receiver with a large $N_{\mathrm{R}}$. Moreover, for quasi-static fading channel, the result of QR factorization that is conducted after channel estimation is available through the entire transmission. Therefore, under quasi-static fading channels, the computational complexity is insensitive to $N_{\mathrm{R}}$.
The quantitative analysis of computational complexity will be proposed in Section VII.

When high data rates are required, the QR-aided ML detection algorithm in Algorithm 2 is still computationally expensive because it has to jointly search all combinations of ${q}_{\mathrm{V}}$, ${q}_{\mathrm{H}}$, and $k$.
Thus, there is a strong demand for computationally efficient detection algorithms that are able to achieve both acceptable ABEP performance and computational complexity. In this section, two computational detection algorithms, i.e., a QR-aided successive interference cancellation (SIC) detection algorithm and a sphere-decoding detection algorithm are proposed.

\begin{algorithm}[!t]
\caption{Pseudo-code for the SIC detection algorithm}
\label{linearsic}
\begin{algorithmic}[1]
\Require $\mathbf{H}$, $\mathbf{y}$
\Ensure $\hat{k}$,$\hat{q}_{\mathrm{V}}$,$\hat{q}_{\mathrm{H}}$
\State $[\mathbf{Q},\mathbf{R}] \gets \mathrm{qr}(\mathbf{H})$, $\mathbf{q}_1 \gets \mathbf{Q}(:,1)^{\mathrm{H}}$, $\mathbf{q}_2 \gets \mathbf{Q}(:,2)^{\mathrm{H}}$, $r_1 \gets \mathbf{R}(1,1)$, $r_{12} \gets \mathbf{R}(1,2)$, $r_2 \gets \mathbf{R}(2,2)$, $\tilde{y}_1 \gets \mathbf{q}_1\mathbf{y}$, $\tilde{y}_2 \gets \mathbf{q}_2\mathbf{y}$
\State $\hat{q}_{\mathrm{H}} \gets \mod\left(\left\lceil \frac{M[\pi u(-r_{2})+\angle(\tilde{y}_2)]}{2\pi}\right\rfloor,M\right)$, $\hat{q}_{\mathrm{V,0}} \gets 0$,  $\hat{k}_{0}\gets1$, $d_{\mathrm{s}} \gets +\infty$
\For {$\hat{k}_{0,\mathrm{SIC}}=1$ to $K$}
\State $\beta\gets \frac{M\left\{ \pi u(-r_1)+\angle\left[\tilde{y}_1-r_{12}\sin\epsilon_{\hat{k}_{0}}
\exp\left(\frac{j2\pi \hat{q}_{\mathrm{H}}}{M}\right)\right]\right\}}{2\pi}$
\State $\hat{q}_{\mathrm{V,0}}\gets \mod\left(\left\lceil \beta\right\rfloor,M\right)$
\State $d_{\mathrm{s}}' \gets
                |\tilde{y}_1-r_1\cos\epsilon_{\hat{k}_{0}}\exp(j\frac{2\pi \hat{q}_{\mathrm{V,0}}}{M})-r_{12}\sin\epsilon_{\hat{k}_{0}}\exp(j\frac{2\pi \hat{q}_{\mathrm{H}}}{M})|^2+|\tilde{y}_2-r_{2}\sin\epsilon_{k_0}\exp(j\frac{2\pi \hat{q}_{\mathrm{H}}}{M})|^2$
\If {$d_{\mathrm{s}}'<d_{\mathrm{s}}$}
\State $\hat{k}\gets \hat{k}_{0}$, $\hat{q}_{\mathrm{V}} \gets \hat{q}_{\mathrm{V,0}}$, $d_{\mathrm{s}}\gets d_{\mathrm{s}}'$
\EndIf
\EndFor
\end{algorithmic}
\end{algorithm}

\subsection{SIC detection algorithm}

In conventional MIMO systems, QR-aided SIC detection algorithms were initially proposed \cite{SIC1,SIC2}. Here, the SIC linear receiver for the PolarSK system is proposed.
Given the QR decomposition of the channel matrix $\mathbf{H}=\mathbf{QR}$ as Eq. (\ref{qrofh}), we obtain

\begin{eqnarray}
&&\label{y1eq}\hspace{-0.5in}\tilde{y}_1= r_1x_{\mathrm{V}}+r_{12}x_{\mathrm{H}}+\mathbf{q}_2\mathbf{w},
\\
&&\label{y2eq}\hspace{-0.5in}\tilde{y}_2= r_2x_{\mathrm{H}}+\mathbf{q}_1\mathbf{w}.
\end{eqnarray}

From (\ref{y2eq}), we observe that $\hat{q}_{\mathrm{H}}$ can be directly obtained by detecting the phase of $\tilde{y}_2$, i.e.,
\begin{eqnarray}
\label{sicqh}
\hat{q}_{\mathrm{H}}=\mathrm{mod}\left(\left\lceil \frac{M[\pi u(-r_{2})+\angle(\tilde{y}_2)]}{2\pi}\right\rfloor,M\right),
\end{eqnarray}
where $\angle(\circ)$ denotes the argument of a complex number, $u(\circ)$ denotes the Heaviside step function that determines whether $r_{2}$ is positive or negative,
$\lceil\circ \rfloor$ rounds a real number to the nearest integer, and $\mathrm{mod}$ denotes the modulus after division.
 
For the PolarSK system, even though $\hat{q}_{\mathrm H}$ can be correctly estimated, the optimum estimation of $\hat{q}_{\mathrm V}$ still depends on $\hat{k}$. Therefore, we can not obtain the optimum $\hat{q}_{\mathrm V}$  directly even after we have obtained $\hat{q}_{\mathrm H}$  as the SIC detection algorithm for conventional MIMO systems. Instead, we need to traverse pairs of $\hat{q}_{\mathrm V}$ and  $\hat{k}$ to find an optimal solution. Fortunately, 
for an arbitrary possible candidate of $\hat{k}_0$, substituting $\hat{q}_{\mathrm{H}}$ into (\ref{y1eq}), we can obtain a closed-form optimum estimation of $\hat{q}_{\mathrm{V},0}(\hat{k}_0)$ as
\begin{eqnarray}
\label{qvsic0}
\hat{q}_{\mathrm{V,0}}(\hat{k}_0)=\mathrm{mod}\left(\left\lceil \alpha_{\mathrm{V}}\right\rfloor,M\right),
\end{eqnarray}
where
\begin{eqnarray}
\begin{array}{l}
\alpha_{\mathrm{V}}\triangleq \frac{M\left\{ \pi u(-r_1)+\angle\left[\tilde{y}_1-r_{12}\sin\epsilon_{\hat{k}_{0}}
\exp\left(\frac{j2\pi \hat{q}_{\mathrm{H}}}{M}\right)\right]\right\}}{2\pi},
\end{array}
\end{eqnarray}
and $\tilde{y}_1-r_{12}\sin\epsilon_{\hat{k}_{0}}
\exp\left(\frac{j2\pi \hat{q}_{\mathrm{H}}}{M}\right)$ is the SIC operation that subtracts the signal with $\hat{q}_{\mathrm{H}}$ from the combined signal $\tilde{y}_1$, then $\hat{q}_{\mathrm{V},0}$ is detected from the residue under $\hat{k}_0$.
Substituting (\ref{qvsic0}) into (\ref{mlck}), we obtain the SIC detection of $\hat{k}$ and $\hat{q}_{\mathrm{V}}$ as
\begin{eqnarray}
&&\label{sick}\hat{k}=
\underset
{\hat{k}_0}{\arg\min}
\left\{
\begin{smallmatrix}
|\tilde{y}_1-r_1\cos\epsilon_{\hat{k}_0}\exp\left(j\frac{2\pi \hat{q}_{\mathrm{V},0}(\hat{k}_0)}{M}\right)\\
-r_{12}\sin\epsilon_{\hat{k}_0}\exp\left(j\frac{2\pi \hat{q}_{\mathrm{H}}}{ M}\right)|^2\\
+|\tilde{y}_2-r_{2}\sin\epsilon_{\hat{k}_0}\exp\left(j\frac{2\pi \hat{q}_{\mathrm{H}}}{ M}\right)|^2
\end{smallmatrix}
\right\},\\
&&\label{sicqv}\hat{q}_{\mathrm{V}}=\hat{q}_{\mathrm{V},0}(\hat{k}).
\end{eqnarray}
Following Eqs. (\ref{sicqh},\ref{sick}-\ref{sicqv}) the pseudo-code for the SIC detection algorithm is proposed in Algorithm 3.

It is still a challenge to analytically evaluate the ABEP performance of the SIC detection algorithm. In this paper, the ABEP performance of the SIC detection algorithm is obtained through Monte Carlo simulations.

\subsection{Optimum sphere-decoding detection algorithm}

\begin{figure*}[t]
\begin{eqnarray}
\setcounter{equation}{44}
\label{reforms2}
{\mathrm{S}_2 =
\left\{\hat{k},\hat{q}_{\mathrm{V}},\hat{q}_{\mathrm{H}}\left|
\begin{array}{l}
 \frac {d^2-|\tilde{y}_1-r_{12}\sin\epsilon_{\hat{k}}\exp(j\frac{2\pi \hat{q}_{\mathrm{H}}}{ M})|^2-|r_1|^2\cos^2\epsilon_{\hat{k}}-|\tilde{y}_2-r_{2}\sin\epsilon_{\hat{k}}\exp(j\frac{2\pi \hat{q}_{\mathrm{H}}}{ M})|^2}{2|\tilde{y}_1-r_{12}\sin\epsilon_{\hat{k}}\exp(j\frac{2\pi \hat{q}_{\mathrm{H}}}{ M})||r_1|\cos\epsilon_{\hat{k}}}
\\
\leq \cos(\pi u(-r_{1})+\angle[\tilde{y}_1-r_{12}\sin\epsilon_{\hat{k}}\exp(j\frac{2\pi \hat{q}_{\mathrm{H}}}{ M})]-\frac{2\pi\hat{q}_{\mathrm{V}}}{M})
\end{array}
\right.\right\}.}
\end{eqnarray}
\hrulefill
\end{figure*}

\begin{algorithm}[!t]
\caption{Pseudo-code for computational SD algorithm.}
\label{sd1}
\begin{algorithmic}[1]
{
\Require $\mathbf{H}$, $\mathbf{y}$
\Ensure $\hat{k}$,$\hat{q}_{\mathrm{V}}$,$\hat{q}_{\mathrm{H}}$
\State $[\mathbf{Q},\mathbf{R}] \gets \mathrm{qr}(\mathbf{H})$, $\mathbf{q}_1 \gets \mathbf{Q}(:,1)^{\mathrm{H}}$, $\mathbf{q}_2 \gets \mathbf{Q}(:,2)^{\mathrm{H}}$, $r_1 \gets \mathbf{R}(1,1)$, $r_{12} \gets \mathbf{R}(1,2)$, $r_2 \gets \mathbf{R}(2,2)$,
$\tilde{y}_1 \gets \mathbf{q}_1\mathbf{y}$, $\tilde{y}_2 \gets \mathbf{q}_2\mathbf{y}$
\State $[\hat{q}_{\mathrm{H,SIC}}, \hat{q}_{\mathrm{V,SIC}}, \hat{k}_{\mathrm{SIC}}] \gets \text{Algorithm \ref{linearsic}}(\mathbf{H},\mathbf{y})$
\State  $d_{\mathrm{s}} \gets
|\tilde{y}_1-r_1\cos\epsilon_{\hat{k}_{\mathrm{SIC}}}\exp(j\frac{2\pi \hat{q}_{\mathrm{V,SIC}}}{M})-r_{12}\sin\epsilon_{\hat{k}_{\mathrm{SIC}}}\exp(j\frac{2\pi \hat{q}_{\mathrm{H,SIC}}}{M})|^2+|\tilde{y}_2-r_{2}\sin\epsilon_{k_0}\exp(j\frac{2\pi \hat{q}_{\mathrm{H,SIC}}}{M})|^2$
\State $\alpha_{\mathrm{H}}\gets \frac{M\left[\pi u(-r_{2})+\angle(\tilde{y}_2)\right]}{2\pi}$
\If {$\lfloor\alpha_{\mathrm{H}}\rfloor==\lceil\alpha_{\mathrm{H}}\rfloor$}
\State $C_{\mathrm{H}}\gets1$
\Else
\State $C_{\mathrm{H}}\gets-1$
\EndIf
\For {$\hat{k}_{0}=1$ to $K$}
\State $\hat{q}_{\mathrm{H,0}}\gets \hat{q}_{\mathrm{H,SIC}}$
\For {$i_{q_{\mathrm{H}}}=0$ to $M-1$}
\State $\hat{q}_{\mathrm{H,0}}\gets \mod(\hat{q}_{\mathrm{H,0}}+C_{\mathrm{H}}(-1)^{i_{q_{\mathrm{H}}}+1}i_{q_{\mathrm{H}}},M)$
\If {$\left|\tilde{y}_2-r_2\sin\epsilon_{\hat{k}_{0}}\exp\left(\frac{j2\pi \hat{q}_{\mathrm{H,0}}}{M}\right)\right|^2>d_{\mathrm{s}}$}
\State BREAK
\EndIf
\State $\alpha_{\mathrm{V}}\gets
\frac{M\left\{ \pi u(-r_1)+\angle\left[\tilde{y}_1-r_{12}\sin\epsilon_{\hat{k}_{0}}
\exp\left(\frac{j2\pi \hat{q}_{\mathrm{H,0}}}{M}\right)\right]\right\}}{2\pi}$
\If {$\lfloor\alpha_{\mathrm{V}}\rfloor==\lceil\alpha_{\mathrm{V}}\rfloor$}
\State $C_{\mathrm{V}}\gets1$
\Else
\State $C_{\mathrm{V}}\gets-1$
\EndIf
\State $\hat{q}_{\mathrm{V,0}}\gets
\mod\left(\left\lceil \alpha_{\mathrm{V}} \right\rfloor,M\right)$
\For {$i_{q_{\mathrm{V}}}=0$ to $M-1$}
\State $\hat{q}_{\mathrm{V,0}}\gets \mod(\hat{q}_{\mathrm{V,0}}+C_{\mathrm{V}}(-1)^{i_{q_{\mathrm{V}}}+1}i_{q_{\mathrm{V}}},M)$
\State $d_{\mathrm{s}}'\gets  |\tilde{y}_1-r_1\cos\epsilon_{\hat{k}_{0}}\exp(j\frac{2\pi \hat{q}_{\mathrm{V,0}}}{M})-r_{12}\sin\epsilon_{\hat{k}_{0}}\exp(j\frac{2\pi \hat{q}_{\mathrm{H,0}}}{M})|^2+|\tilde{y}_2-r_{2}\sin\epsilon_{k_0}\exp(j\frac{2\pi \hat{q}_{\mathrm{H,0}}}{M})|^2$
\If {$d_{\mathrm{s}}'>d_{\mathrm{s}}$}
\State BREAK
\EndIf
\State $\hat{k}\gets \hat{k}_{0}$, $\hat{q}_{\mathrm{H}} \gets \hat{q}_{\mathrm{H,0}}$, $\hat{q}_{\mathrm{V}} \gets \hat{q}_{\mathrm{V,0}}$
\EndFor
\EndFor
\EndFor
}
\end{algorithmic}
\end{algorithm}

\begin{figure} [!t]
\centering
\includegraphics[width=3.5in]{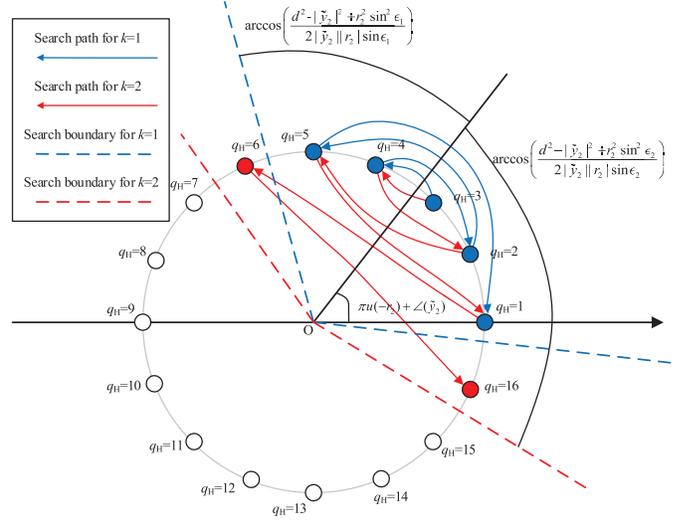}
\caption{Geometry of  $\mathrm{S}_1$ with $K=2$ and $M=16$. Blue-colored points represent those points which are inside $\mathrm{S}_1$ with arbitrary $\hat{k}$. Red-colored points represent those points which are inside $\mathrm{S}_1$ only with $\hat{k}=2$. For the SIC receiver, $\hat{q}_{\mathrm{H}}=3$.
}\label{s1fig}
  \vspace{-10pt}
\end{figure}

It is predictable that compared with the optimum ML receiver, a main drawback of the computational SIC detection algorithm is the loss of ABEP performance.
Whereas the optimum QR-ML detection algorithm involves exhaustive search to find the optimum solution, which has to consider all the legitimate symbol candidates of the number of $KM^2$. In this subsection, we introduce a computational optimum SD detection algorithm, which makes decisions on each possible node only and, as consequence, cuts some node to reduce the number of visited nodes without loss of ABEP performance. The basic idea of the SD detection algorithm is to search over only constellation points which lie in a certain sphere of radius $d$ around the received vector $\left[\begin{smallmatrix}
\tilde{y}_1\\
\tilde{y}_2
\end{smallmatrix}\right]$, i.e.,
\begin{eqnarray}
\setcounter{equation}{38}
\label{sphere0}
\mathrm{S}=
\left\{\hat{k},\hat{q}_{\mathrm{V}},\hat{q}_{\mathrm{H}}\left|
\begin{smallmatrix}
|\tilde{y}_1-r_1\cos\epsilon_{\hat{k}}\exp\left(j\frac{2\pi \hat{q}_{\mathrm{V}}}{ M}\right)\\
-r_{12}\sin\epsilon_{\hat{k}}\exp\left(j\frac{2\pi \hat{q}_{\mathrm{H}}}{ M}\right)|^2\\
+|\tilde{y}_2-r_{2}\sin\epsilon_{\hat{k}}\exp\left(j\frac{2\pi \hat{q}_{\mathrm{H}}}{ M}\right)|^2
\leq d^2
\end{smallmatrix}\right.\right\},
\end{eqnarray}%
to reduce the search space \cite{SD1,SD2}. The decoding process is a decision on a tree with 4 layers. The first layer only have one root node. In the second layer, $K$ branches depart from the root node corresponding to all possible $k$ and each node in the second layer has $M$ branch nodes corresponding to all possible $q_{\mathrm{H}}$ in the third layer. Each branch node in the third layer has $M$ leaf nodes corresponding to $q_{\mathrm{V}}$ in the forth layer. The pseudo-code of the SD detection algorithm is shown in Algorithm \ref{sd1}, where  $\lceil\circ\rceil$ rounds a real number to the nearest larger integer, and $\lfloor\circ\rfloor$ rounds a real number to the nearest smaller integer. The explanation of and is in the following context.

Firstly, it is necessary to find a principle with low computational complexity to determine which constellation points are inside the sphere (\ref{sphere0}). By respectively defining spheres $\mathrm{S}_1$ and $\mathrm{S}_2$ as 

\begin{eqnarray}
\setcounter{equation}{39}
\label{s1}
{\mathrm{S}_1 \triangleq
\left\{\hat{k},\hat{q}_{\mathrm{V}},\hat{q}_{\mathrm{H}}\left|\begin{array}{l}
|\tilde{y}_2-r_{2}\sin\epsilon_{\hat{k}}\exp(j\frac{2\pi \hat{q}_{\mathrm{H}}}{ M})|^2
\leq d^2\end{array}
\right.\right\}},
\end{eqnarray}
and
\begin{eqnarray}
\label{s2}
{\mathrm{S}_2 \triangleq
\left\{\hat{k},\hat{q}_{\mathrm{V}},\hat{q}_{\mathrm{H}}\left|
\begin{array}{l}
|\tilde{y}_1-r_1\cos\epsilon_{\hat{k}}\exp(j\frac{2\pi \hat{q}_{\mathrm{V}}}{ M})
\\-r_{12}\sin\epsilon_{\hat{k}}\exp(j\frac{2\pi \hat{q}_{\mathrm{H}}}{ M})|^2
\leq d^2
\\-|\tilde{y}_2-r_{2}\sin\epsilon_{\hat{k}}\exp(j\frac{2\pi \hat{q}_{\mathrm{H}}}{ M})|^2
\end{array}
\right.\right\}},
\end{eqnarray}
Eq. (\ref{sphere0}) is rewritten as
\begin{eqnarray}
\setcounter{equation}{41}
\mathrm{S}=\mathrm{S}_1\cap \mathrm{S}_2.
\end{eqnarray}

Reforming the sphere $\mathrm{S}_1$ and following some simple algebraic manipulations, we obtain
\begin{eqnarray}
\label{reforms1}
\mathrm{S}_1 =
\left\{\hat{k},\hat{q}_{\mathrm{V}},\hat{q}_{\mathrm{H}}\left|
\begin{array}{l}
\frac{d^2-|\tilde{y}_2|^2-r_{2}^2\sin^2\epsilon_{\hat{k}}}{2|\tilde{y}_2||r_{2}|\sin\epsilon_{\hat{k}}}
\leq \\ \cos\left[\pi u(-r_{2})+\angle(\tilde{y}_2)-\frac{2\pi\hat{q}_{\mathrm{H}}}{M}\right]
\end{array}
\right.\right\}.
\end{eqnarray}

From (\ref{reforms1}), we observe that the right hand side of the inequality decreases with the increasing distance of the angle $\pi u(-r_{2})+\angle(\tilde{y}_2)$ and the angle $\frac{2\pi\hat{q}_{\mathrm{H}}}{M}$. A geometrical illustration of $\mathrm{S}_1$ with $K=2$ and $M=16$ is shown in Fig. \ref{s1fig}. 
Unlike the SD receiver in the conventional TIMO system, which only searches optimal $\hat{q}_{\mathrm{V}}$ and $\hat{q}_{\mathrm{H}}$, the search for optimum $\hat{k}$ is also required for the PolarSK signal detection. Since it is difficult to find a computational approach to determine whether $\hat{k}$ lies in a certain sphere under a given  $\hat{q}_{\mathrm{H}}$ or $\hat{q}_{\mathrm{V}}$, we conduct the SD algorithm to detect  $\hat{q}_{\mathrm{V}}$  for every possible candidate of $\hat{k}$ under given $\hat{q}_{\mathrm{H}}$, and obtain the optimum $\hat{k}$ by the MED principle.
Denoting $\hat{k}_0$ as a possible candidate of $\hat{k}$, the red and the blue dash lines are search boundaries respectively for $\hat{k}_0=1$ and $\hat{k}_0=2$, beyond which notes are out of sphere S$_1$, and therefore the search can be terminated safely. When $d$ gets smaller, the red and the blue dash lines are closer to the solid line and there will be less candidate $\hat{q}_{\mathrm{H}}$.

Then, we clearly observe from Fig. \ref{s1fig} that without considering $\mathrm{S}_2$, the detection of $q_{\mathrm{H},0}$ is the nearest point to the line $\exp[j\pi u(-r_{2})+j\angle(\tilde{y}_2)]$, i.e., $\hat{q}_{\mathrm{H},0}=\mod\left(\left\lceil \frac{M[\pi u(-r_{2})+\angle(\tilde{y}_2)]}{2\pi}\right\rfloor,M\right)$, as given in line (1) of Algorithm \ref{sd1}.
Assuming that $\hat{q}_{\mathrm{H},0}$ is at the clockwise side of $\pi u(-r_{2})+\angle(\tilde{y}_2)$, and inspired by the Schnorr-Euchner strategy \cite{pohst}, the sequence
\begin{eqnarray}
\setcounter{equation}{43}
\hat{q}_{\mathrm{H},0},\hat{q}_{\mathrm{H},0}+1,\hat{q}_{\mathrm{H},0}-2,\hat{q}_{\mathrm{H},0}+3,\hat{q}_{\mathrm{H},0}+M-1
\end{eqnarray}
orders the possible $\hat{q}_{\mathrm{H}}$ in Fig. \ref{s1fig} according to nondecreasing distance from $\hat{q}_{\mathrm{H}}$ to $\pi u(-r_{2})+\angle(\tilde{y}_2)$. A trivial counterpart holds when $\hat{q}_{\mathrm{H},0}$ is at the counter-clockwise side of $\pi u(-r_{2})+\angle(\tilde{y}_2)$. The search following the sequence can safely be terminated as soon as reaching the search boundary, i.e.,  $
|\tilde{y}_2-r_{2}\sin\epsilon_{\hat{k}}\exp(j\frac{2\pi \hat{q}_{\mathrm{H}}}{ M})|^2>d^2$ as in the lines (14-16) of Algorithm \ref{sd1}.
Whether $\hat{q}_{\mathrm{H},0}$ is at the clockwise side of $\pi u(-r_{2})+\angle(\tilde{y}_2)$ is determined in lines (5-9) of Algorithm \ref{sd1}.

Following the idea of reforming $\mathrm{S}_1$, the sphere $\mathrm{S}_2$ is reformed as (\ref{reforms2}), from which the search boundary is obtained for each pair of $\hat{k}_0$ and $\hat{q}_{\mathrm{H},0}$, which respectively denote possible candidates of $\hat{k}$ and $\hat{q}_{\mathrm{H}}$. Given $\hat{k}_0$ and $\hat{q}_{\mathrm{H},0}$, whether $\hat{q}_{\mathrm{V},0}$ is at the clockwise side of $\pi u(-r_1)+\angle\left[\tilde{y}_1-r_{12}\sin\epsilon_{\hat{k}_{0}}
\exp\left(\frac{j2\pi \hat{q}_{\mathrm{H,0}}}{M}\right)\right]$ is firstly determined in lines (18-22) of Algorithm \ref{sd1}.
Then, the search steps of $\hat{q}_{\mathrm{V}}$ are given in lines (23-31) of Algorithm \ref{sd1}. For each pair of $\hat{k}_0$ and $\hat{q}_{\mathrm{H},0}$, the search is terminated until the inequality in Eq. (\ref{reforms2}) does not hold.

\begin{remark}
\label{remak3}
Because each node which is possible to be the optimum one is visited in the SD detection algorithm, the proposed SD detection algorithm has the same detection result as the optimum ML receiver. Therefore, ABEPs of the ML and the SD receivers are identical.
\end{remark}

\section{Numerical results}

In this section, we illustrate the performance of the proposed PolarSK systems via numerical results.

\begin{figure} [!t]
\centering
\subfigure[C$_1$]{\includegraphics[width=1.5in]{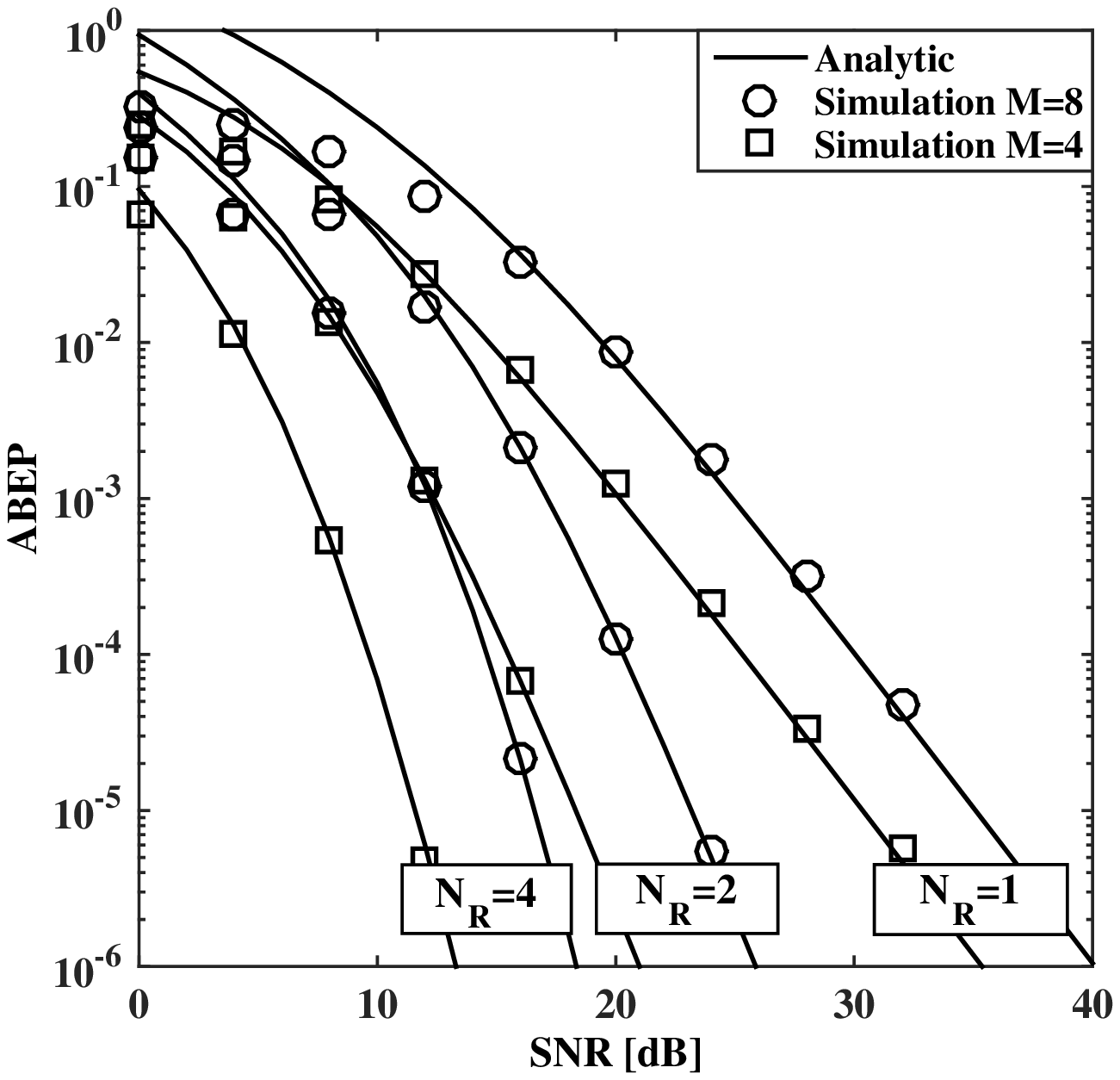}}
\subfigure[C$_2$]{\includegraphics[width=1.5in]{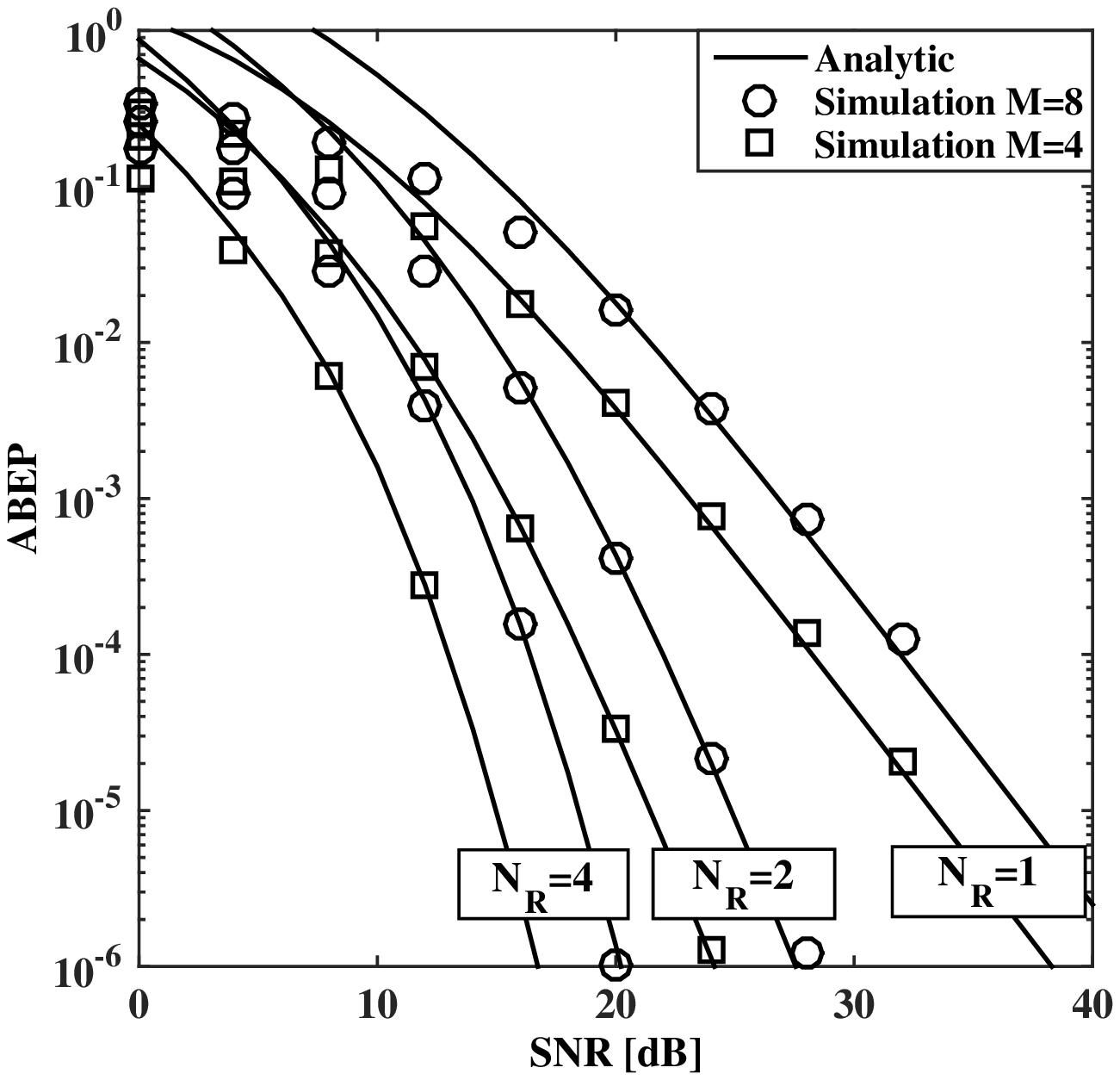}}
\caption{{Verification of analytic ABEPs when $X=-4.5665\mathrm{dB}$.
Markers show Monte Carlo simulations based on $2\times10^6$ realizations, and solid lines and dashed lines show the analytic bounds.
}}\label{C1ABEPverification}
  \vspace{-10pt}
\end{figure}

\begin{figure}[!t]
\centering
\subfigure[$N_{\mathrm{R}}=1$]{\includegraphics[width=1.5in]{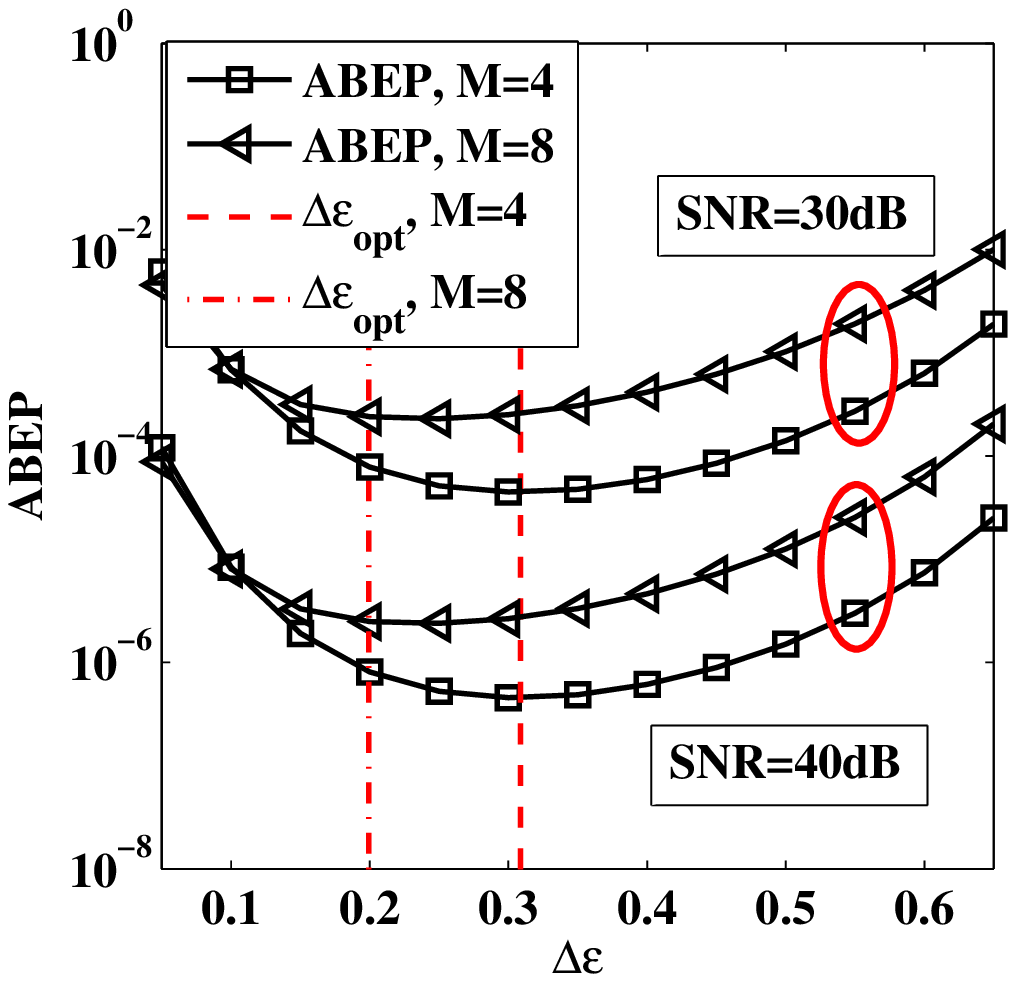}}
\subfigure[$N_{\mathrm{R}}=2$]{\includegraphics[width=1.5in]{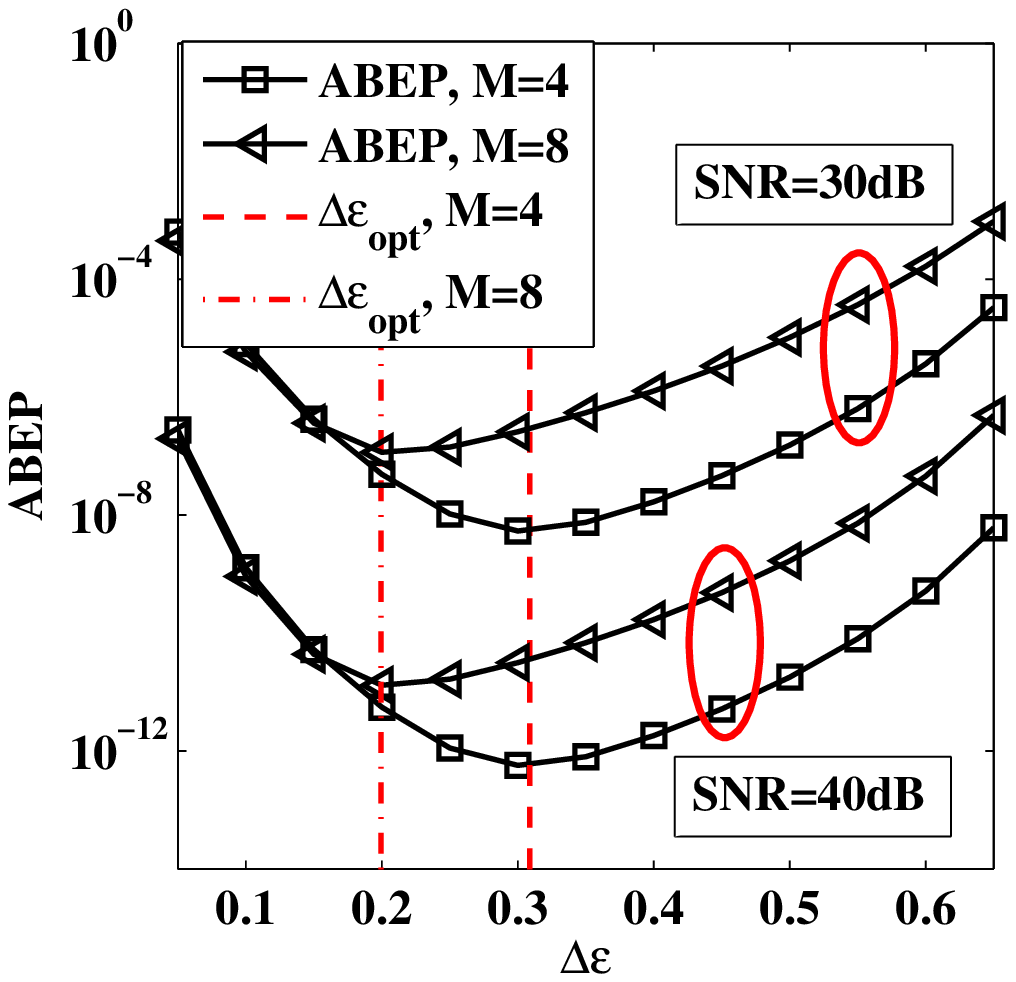}}
\caption{Verification of the optimized $\Delta\epsilon_{\mathrm{opt}}$ for the optimum receiver. C$_2$ signal constellation for $X=-4.5665\mathrm{dB}$. ABEPs of the optimum detector are computed by Eq. (\ref{ABEPeqst8}).
}\label{E2verification}
  \vspace{-10pt}
\end{figure}

\begin{figure}[!t]
\centering
\subfigure[$N_{\mathrm{R}}=1$]{\includegraphics[width=1.5in]{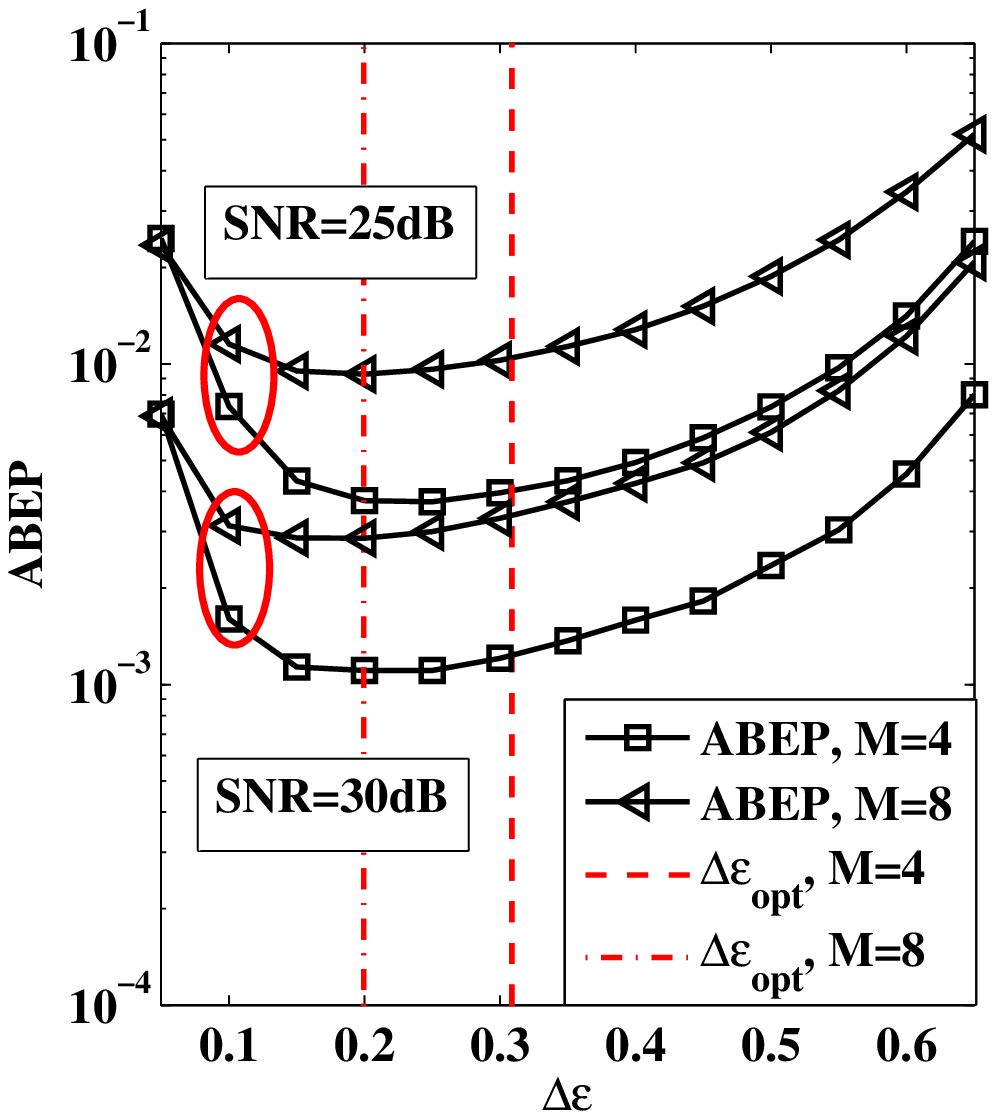}}
\subfigure[$N_{\mathrm{R}}=2$]{\includegraphics[width=1.5in]{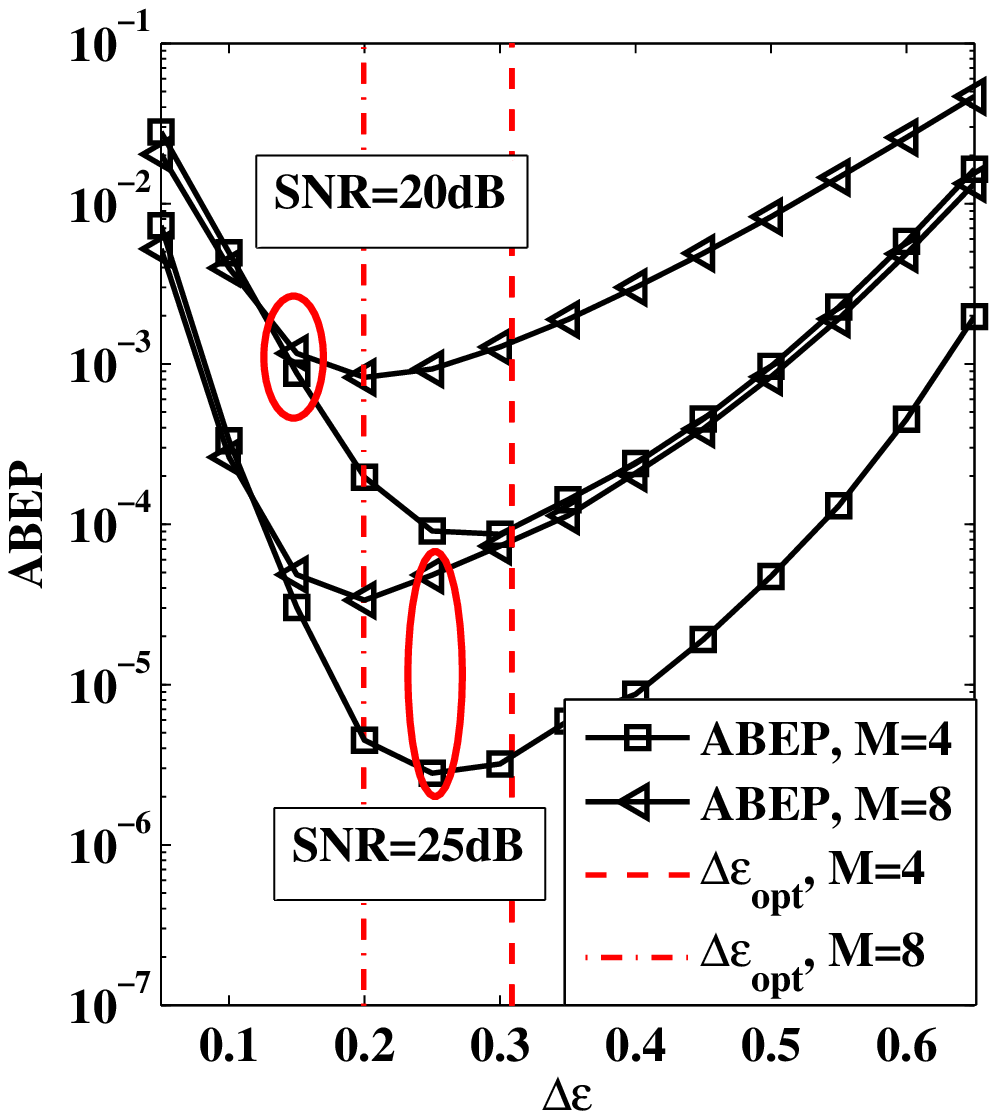}}
\caption{{Verification of the optimized $\Delta\epsilon_{\mathrm{opt}}$ for the linear receiver. C$_2$ signal constellation for $X=-4.5665\mathrm{dB}$. ABEPs of the SIC detection algorithm are obtained by Monte Carlo simulations based on $2\times10^6$ realizations.}
}\label{E2verificationlinear}
  \vspace{-10pt}
\end{figure}

\subsection{Validation of analytic results}

We start by illustrating relevant numerical examples with the aim of validating the accuracy of the proposed mathematical frameworks.

\subsubsection{ABEP}

In Fig. \ref{C1ABEPverification}, for C$_1$ and  C$_2$ PolarSK systems the analytic upper bounds on the ABEP and Monte Carlo simulation results for $2\times10^6$ channel realizations are plotted when $X=4.5665\mathrm{dB}$.
The upper bound on the ABEP is computed by (\ref{ABEPeqst8}). It is observed that the analytic ABEP well overlap with Monte Carlo simulations in the high SNR regime.

\subsubsection{Optimum $\epsilon_k$}

In sections IV, we derived the optimum $\epsilon_k$ in terms of ABEP of optimum receiver. In order to investigate the validity of Algorithm 1, Fig. \ref{E2verification} plots the ABEP of C$_2$ PolarSK behave as a function of $\Delta \epsilon$ using the analytic results in (\ref{ABEPeqst8}) and (\ref{finaloptimal}). We observe that the PolarSK scheme yields the lowest ABEP when $\Delta \epsilon=\Delta \epsilon_{\mathrm{opt}}$.

However, the optimum $\epsilon_k$ is obtained on the basis of the optimum receiver. It is required to investigate if it also works for the SIC detection algorithm. In Fig. \ref{E2verificationlinear}, the ABEP of C$_2$ PolarSK using SIC detection algorithm against $\Delta \epsilon$ is plotted through Monte Carlo simulations based on $2\times 10^6$ realizations. It is observed that the optimum $\epsilon_k$ works well for the SIC detection algorithm. Therefore, we conclude that the optimal $\Delta\epsilon_{\mathrm{opt}}$ can indeed minimize the ABEP of PolarSK systems for all detection algorithms in this paper.

\begin{figure} [!t]
\centering
\subfigure[C$_1$, $M=4$]{\includegraphics[width=1.5in]{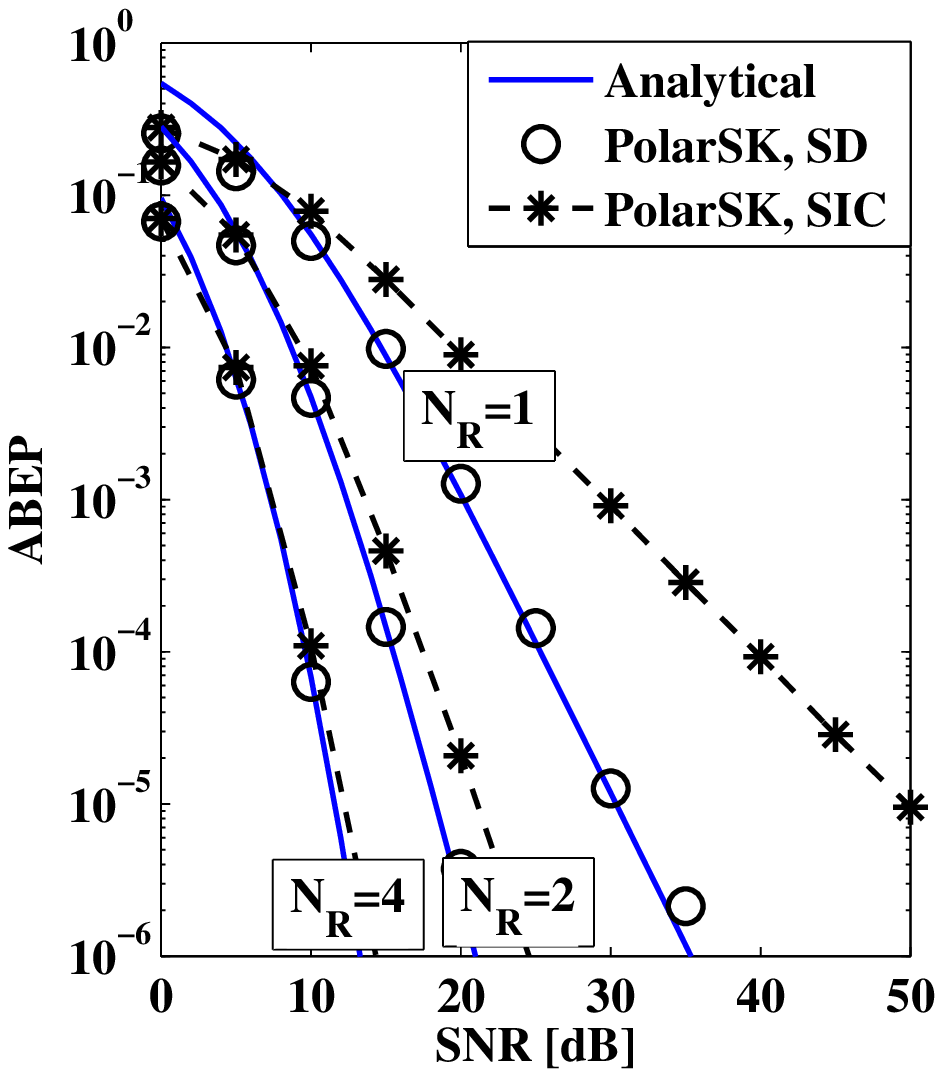}}
\subfigure[C$_2$, $M=4$]{\includegraphics[width=1.5in]{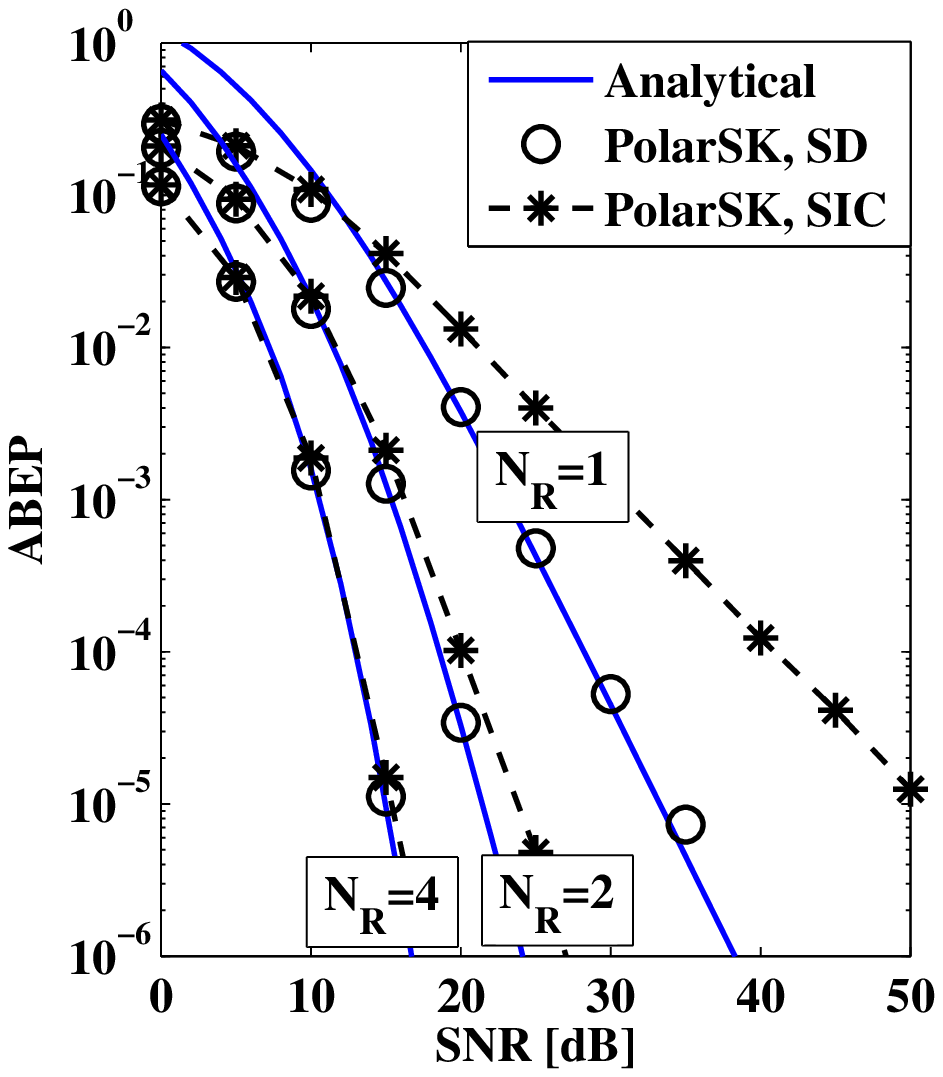}}
\caption{{Comparison of ABEPs of the SIC and the SD detection algorithms.}
}\label{SDvsSIC}
  \vspace{-10pt}
\end{figure}

\subsection{Comparison of receivers}
The computational SIC algorithm proposed in Section V will decrease the computational complexity, whereas the drawback against the SD algorithm is the decreasing of ABEP. In this subsection, ABEPs and computational complexities of the SD and the SIC algorithms are investigated.

\subsubsection{ABEP}
ABEP performances of proposed detection algorithms are shown in Fig. \ref{SDvsSIC} for $M=4$. Since the SD algorithm and the QR-aided ML algorithm have the same ABEP as explained in Remark \ref{remak3}, we do not plot the ABEP of the ML receiver for visibility. From Fig. \ref{SDvsSIC}, the following observations are made.

\begin{enumerate}
\item It is observed that the SIC algorithm can not achieve full diversity, i.e.,  the ABEP performance loss between the ML and the SIC algorithms grows as the SNR.
\item The ABEP performance of the SIC algorithm relies on the orthogonality of the polarized channel significantly. Since the polarized channel is more orthogonal under a larger $N_{\mathrm{R}}$, the SNR gap between ABEPs of the SD and the SIC algorithms decreases with an increasing $N_{\mathrm{R}}$.
\item The SNR loss is small for $N_{\mathrm{R}}>1$ and practical values of SNR. It is nearly 3dB and 1dB for $N_{\mathrm{R}}=2$ and $N_{\mathrm{R}}=4$, respectively. Therefore, the linear SIC receiver is recommended to reduce computational complexity for $N_{\mathrm{R}}>1$, unless the priority of the ABEP performance is higher than the computational complexity.
\item Whereas for $N_{\mathrm{R}}=1$, the ABEP performance of the SIC algorithm is dramatically worse than that of the SD algorithm. For $ABEP=10^{-5}$, the SNR gap between  ABEPs of the SD and the SIC algorithms is even nearly 20dB. Therefore, when $N_{\mathrm{R}}=1$, to obtain an acceptable ABEP performance, the optimum SD algorithm is recommended.
\end{enumerate}

\begin{figure} [!t]
\centering
\subfigure[C$_1$, $M=4$]{\includegraphics[width=1.5in]{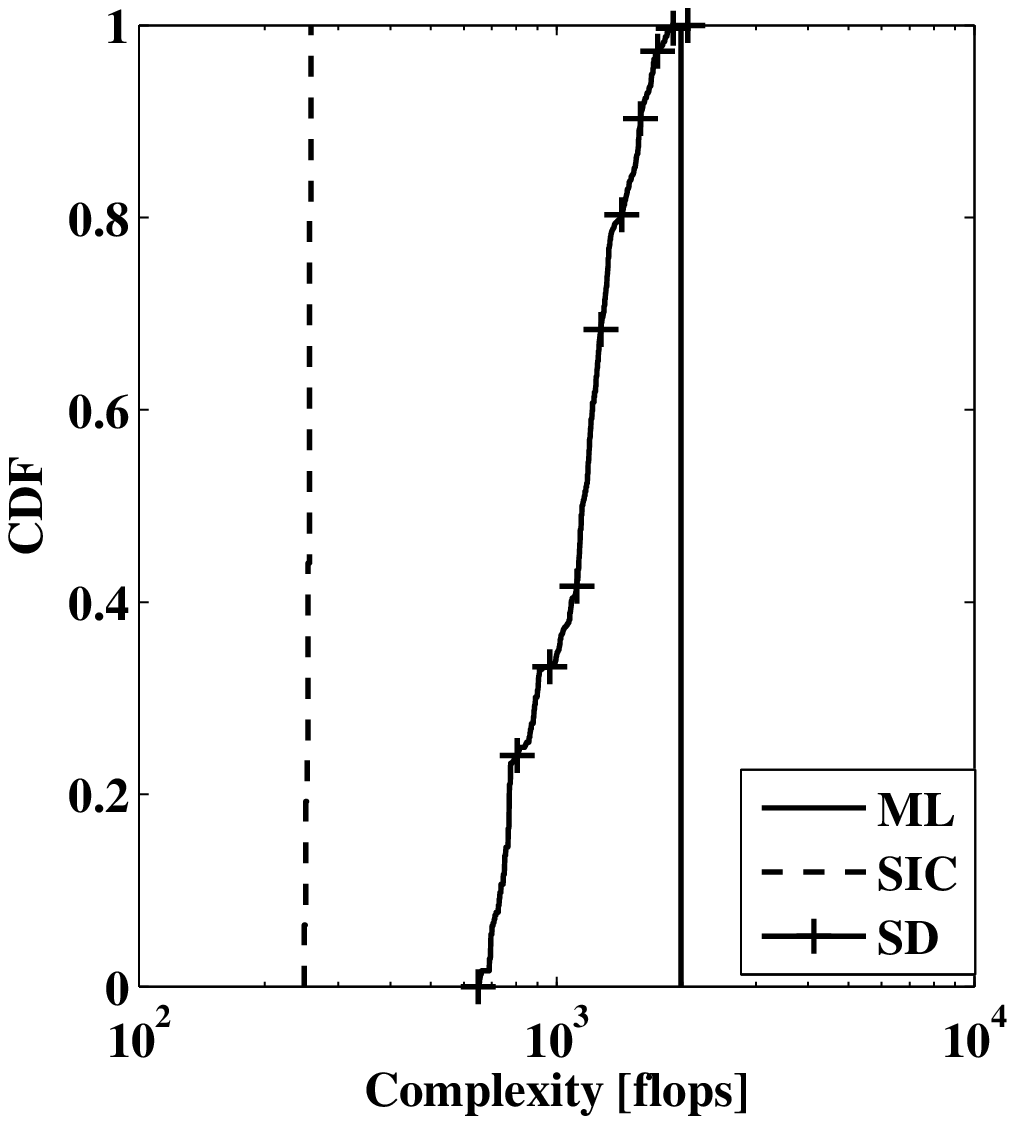}}
\subfigure[C$_1$, $M=8$]{\includegraphics[width=1.5in]{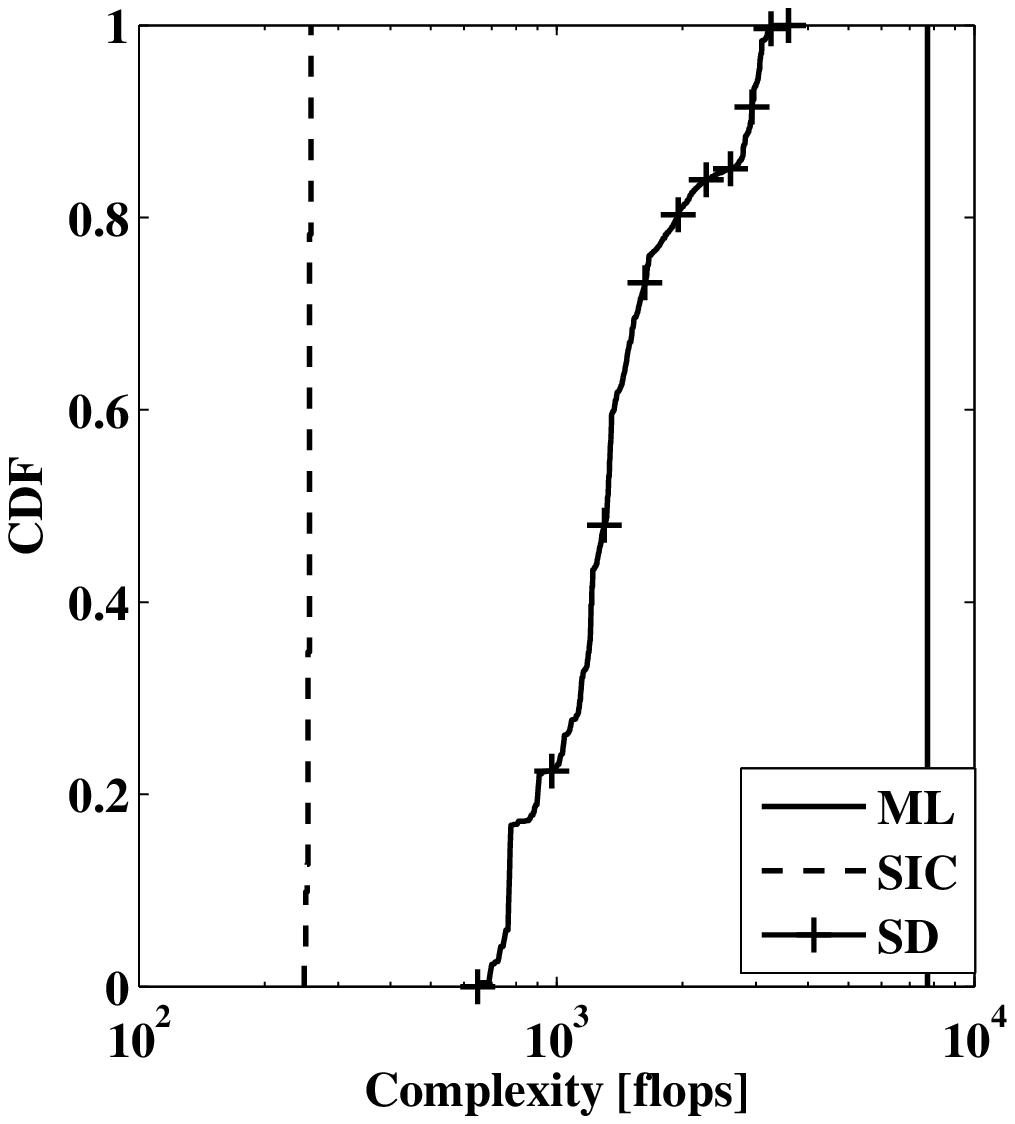}}
\subfigure[C$_2$, $M=4$]{\includegraphics[width=1.5in]{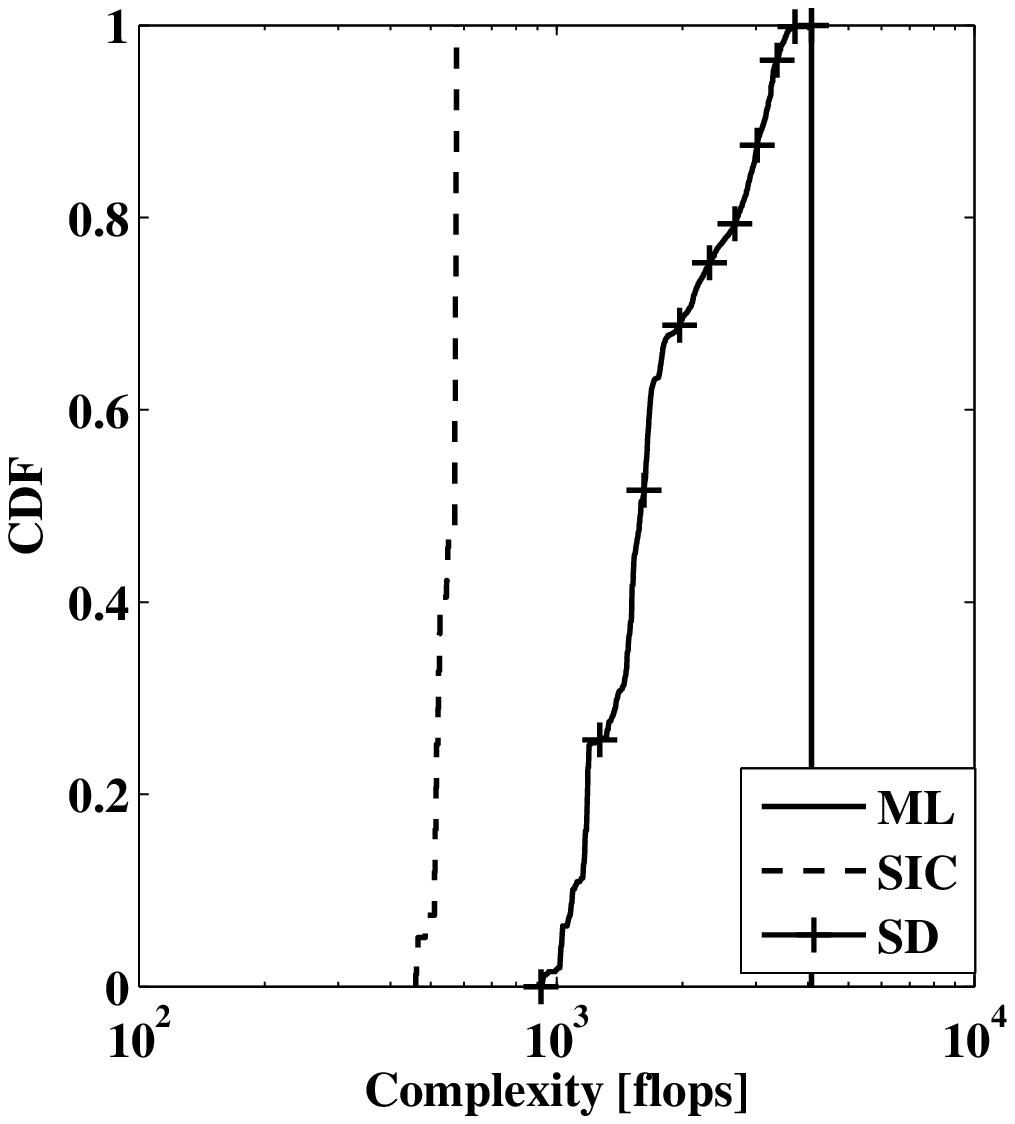}}
\subfigure[C$_2$, $M=8$]{\includegraphics[width=1.5in]{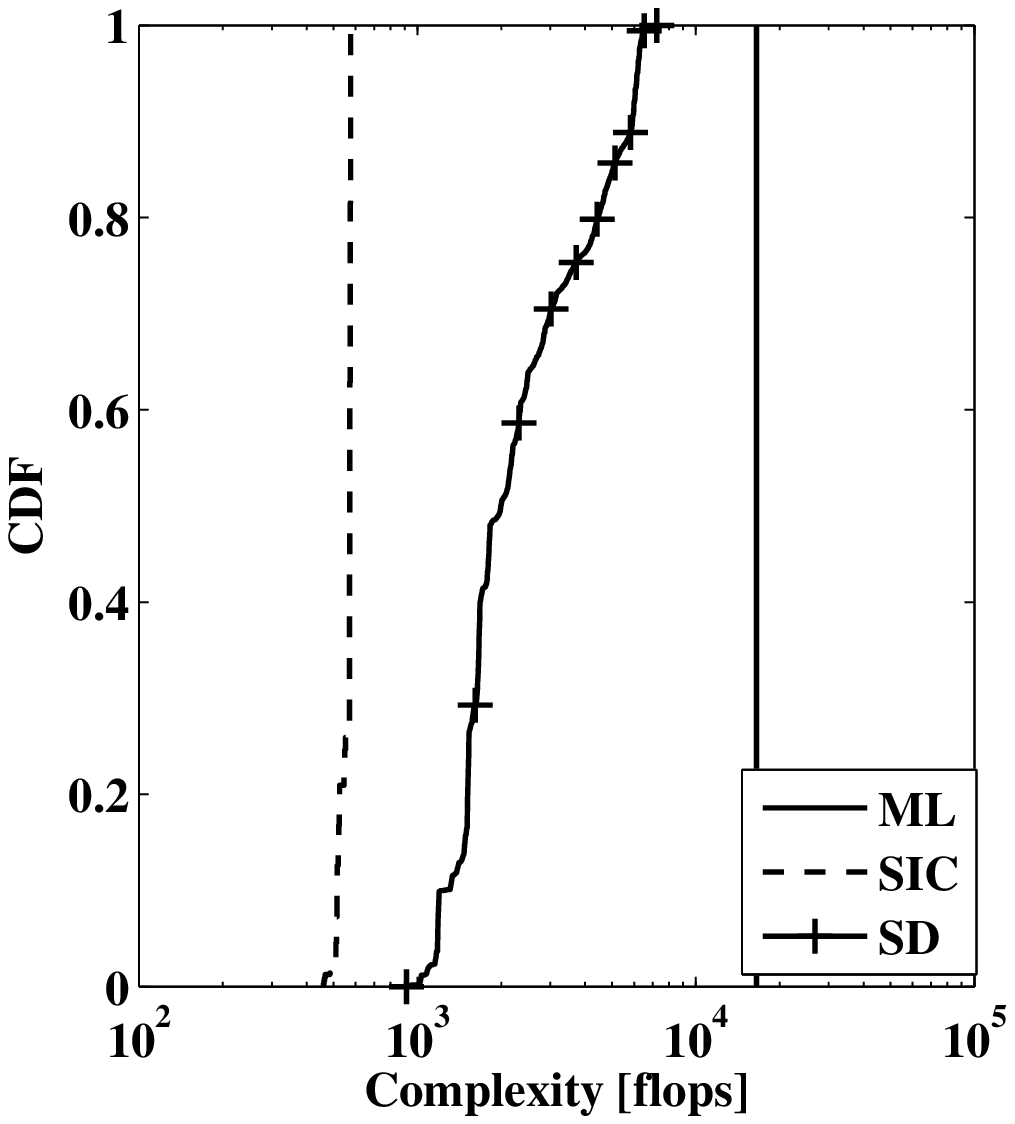}}
\caption{{CDFs of complexity in terms of flops for $N_{\mathrm{R}}=1$, $X=-4.5665$dB and $\rho=20$dB.}
}\label{flops}
  \vspace{-10pt}
\end{figure}

\subsubsection{Computational complexity}

\begin{figure} [!t]
\centering
\subfigure[C$_1$, $M=8$]{\includegraphics[width=1.5in]{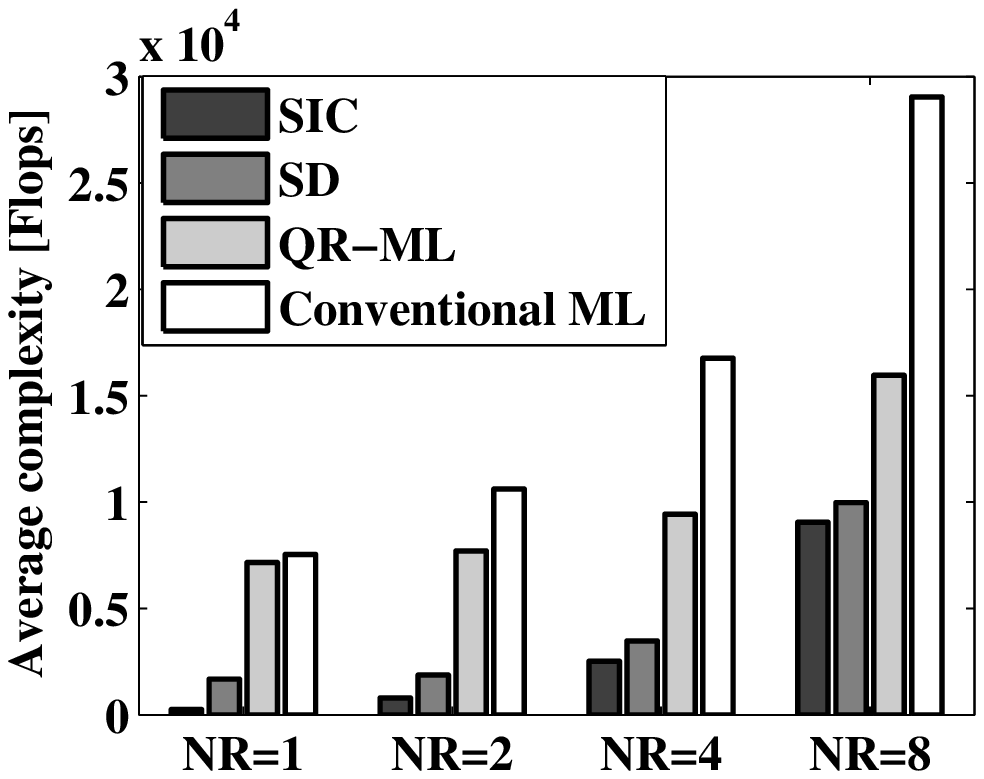}}
\subfigure[C$_2$, $M=8$]{\includegraphics[width=1.5in]{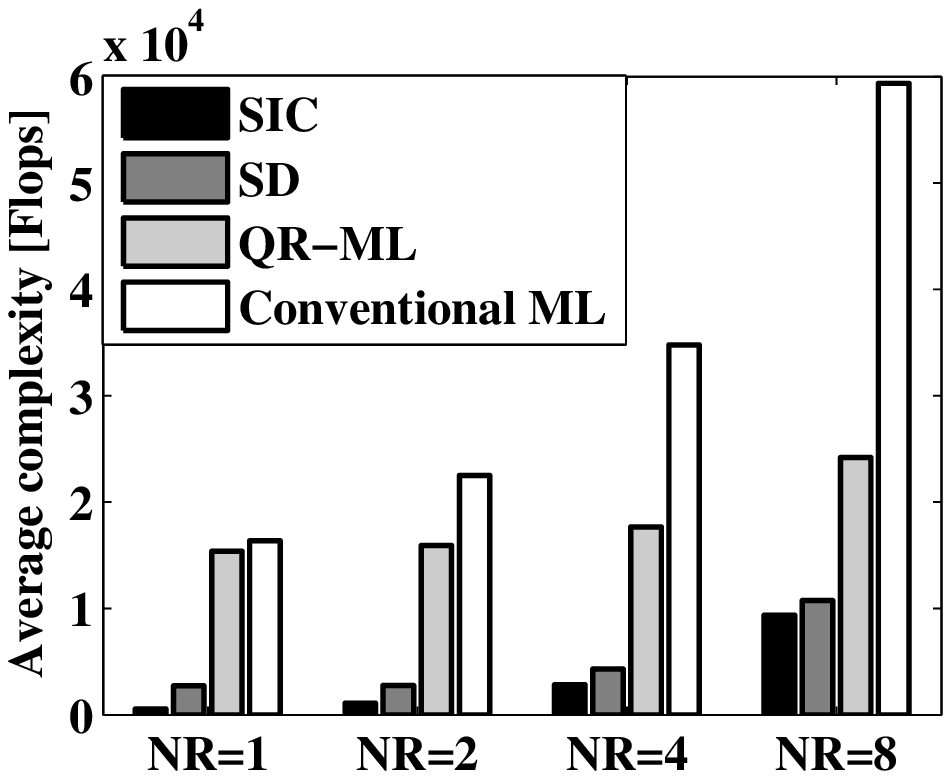}}
\caption{{Average complexity in terms of flops against $N_\mathrm{R}$, $X=-4.5665$dB and $\rho=20$dB. The SIC, the SD, the QR-ML and the conventional ML algorithms are given in Algorithm 3, Algorithm 4, Algorithm 2, and (\ref{mlreceiver1}), respectively.}
}\label{averageflops}
  \vspace{-10pt}
\end{figure}

Here, we quantitatively compare the computational complexities of proposed detection algorithms for PolarSK systems. The computation complexity in terms of floating point operations per second (flops) of the ML, the SD and the SIC algorithms is plotted in Fig. \ref{flops}.
For a fair comparison, we assume that a complex addition requires 2 flops, and a complex multiplication requires 6 flops. The SIC, the SD, and the QR-ML algorithms are respectively given in Algorithm 3, Algorithm 4, and Algorithm 2.
It is observed that the proposed SD algorithm outperforms the QR-ML algorithm because of the reduced number of visited nodes even though the SD and the QR-ML algorithms has the same ABEP. Moreover, unlike SIC and ML algorithms, the computational complexity of the SD receiver varies under different realizations of channel matrix.
In Fig. \ref{averageflops}, average complexities in terms of flops for $M=8$, $X=-4.5665$dB and $\rho=20$dB are plotted, where the conventional ML receiver is given in Eq. (\ref{mlreceiver1}). It is observed from Fig. \ref{averageflops} that

\begin{enumerate}
\item Compared with the conventional ML receiver, the complexity of the QR-aided detection algorithm is not sensitive to the number of receive antennas. For proposed SIC and SD algorithms under fast fading channels, the computational complexity increases with an increasing $N_{\mathrm{R}}$.
\item Typically, small mobile devices can accommodate only one DP antenna. While $N_{\mathrm{R}}=1$, average complexities of the SIC, the SD, and the QR-aided ML algorithms are respectively 255flops, 1678flops and 7154flops for C$_1$ constellation. For C$_2$ constellation, they are 559flops, 2733flops and 15388flops, respectively. We conclude that for $N_{\mathrm{R}}=1$, proposed SIC and SD algorithms reduce the complexity significantly.
\item Since lines (2-10) in Algorithm 3 and lines (3-33) in Algorithm 4 are irrelevant to $N_{\mathrm{R}}$, the increased complexity is mainly contributed by the QR decomposition of the channel matrix. Therefore, we conclude that complexities of SIC and SD receivers with a high $N_{\mathrm{R}}$ under quasi-static fading channels are nearly equal to those of SIC and SD receivers with $N_{\mathrm{R}}=1$ under fast fading channels.
\end{enumerate}

Jointly considering ABEP and computational complexity, we conclude this section with the following design remark.
\begin{remark}
\label{remcofig}
When $N_{\mathrm{R}}>1$, the SNR gap between ABEPs of SD and SIC algorithms is small under practical values of SNR and therefore the computational SIC receiver is recommended.
When $N_{\mathrm{R}}=1$, compared with the SD algorithm, the SIC algorithm is not competitive in terms of ABEP. Since the average complexity of the SD algorithm is only less than 3000 flops for $M=8$, the SD receiver is recommended.
\end{remark}

\begin{figure} [!t]
\centering
\subfigure[{$N_R=1,\rho=30\mathrm{dB}$} ]{\includegraphics[width=1.5in]{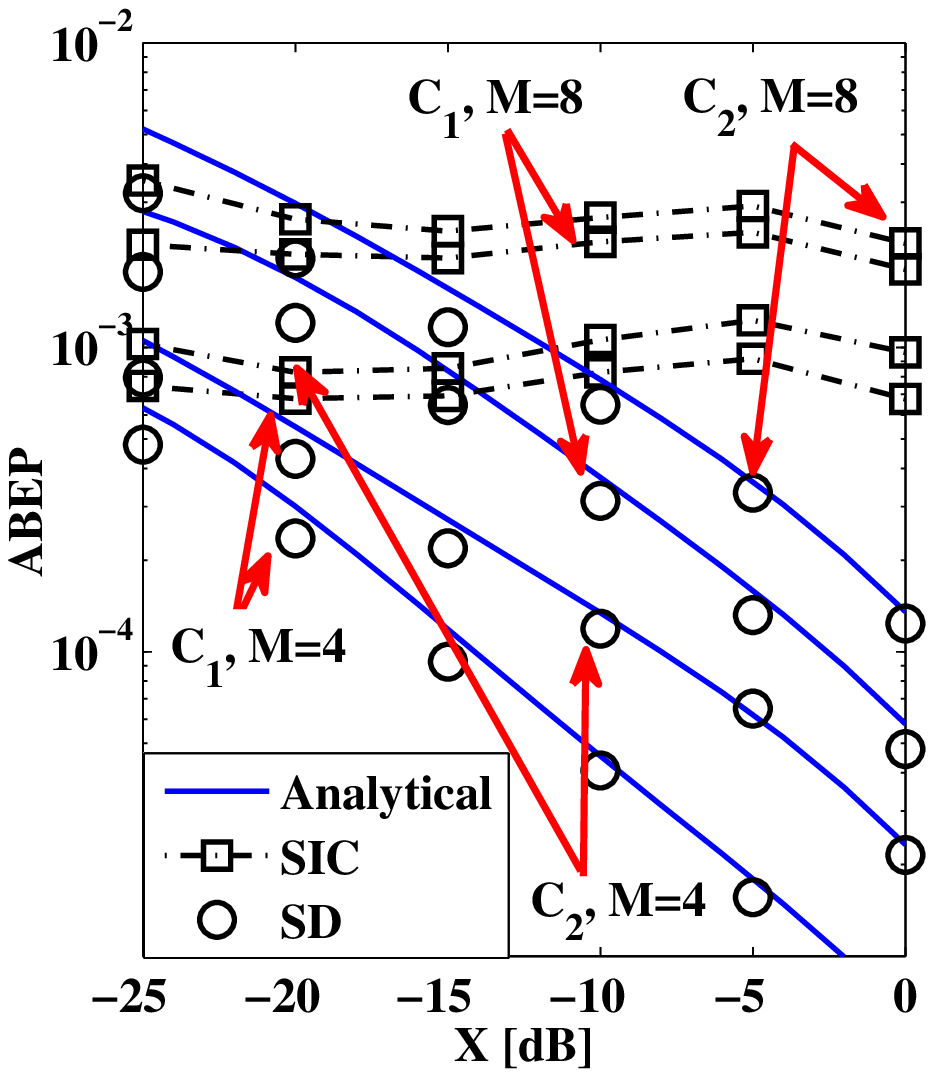}}
\subfigure[{$N_R=2,\rho=20\mathrm{dB}$} ]{\includegraphics[width=1.5in]{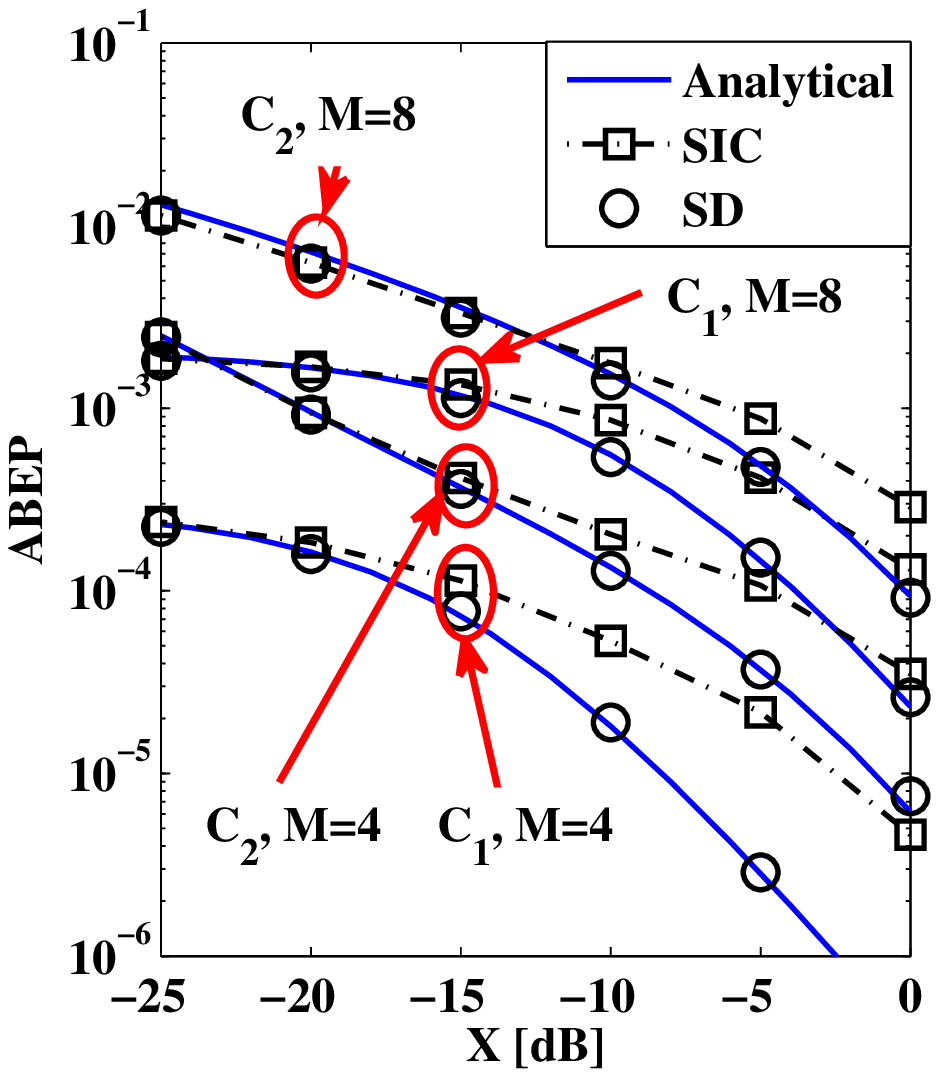}}
\caption{{ABEP of PolarSK systems against $X$. It is observed that the ABEP of the optimal SD detection algorithm deceases with an increasing $X$.}
}\label{VSX}
  \vspace{-10pt}
\end{figure}

\begin{figure}[!t]
\centering
\includegraphics[width=2in]{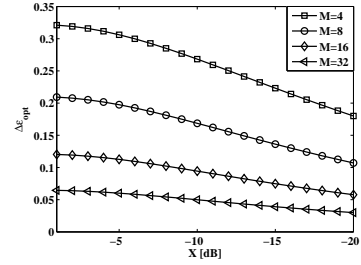}
\caption{$\Delta\epsilon_{\mathrm{opt}}$ against $X$ for C$_2$ constellation.
}\label{E2optimal}
  \vspace{-10pt}
\end{figure}

\subsection{System performance against $X$}
In order to analyze the impact of $X$ on the performance of PolarSK systems, the ABEPs against $X$ are plotted in Fig. \ref{VSX}, from which the following observations are made.
\begin{enumerate}
\item It is observed that the ABEP of the optimal SD algorithm deceases with an increasing $X$. From Eqs. (\ref{lambda1}) and (\ref{lambda2}) we can see that $\Lambda_{\mathrm{V}}$ and $\Lambda_{\mathrm{H}}$ increase with $X$. Since the PEP is proportional to $(\Lambda_{\mathrm{V}}\Lambda_{\mathrm{H}})^{-N_{\mathrm{R}}}$, the PolarSK system can achieve a better performance under a higher $X$.
\item While $X=0$, the polarized channel are fully orthogonality and the SD and the SIC algorithms have the same ABEP performance. As $X$ decreases, the difference between ABEPs of SD and SIC receivers becomes negligible. Therefore, under a propagation scenario with a low $X$, the SIC algorithm, which has a lower computational complexity than than the SD algorithm, is recommended.
\item On one hand, while $X$ increases, the polarized channel has more orthogonality for $N_{\mathrm{R}}=1$, which enhances the ABEP performance of the SIC algorithm. On the other hand, under a fixed orthogonality, with an increasing $X$, the SIC receiver has more received energy for each symbol. Therefore, $X$ has no monotonicity on the ABEP of the SIC algorithm while $N_{\mathrm{R}}=1$.
\item While $N_{\mathrm{R}}=2$, the polarized channel is quasi-orthogonal although $X$ is high, and the impact of $X$ on the received energy dominates the ABEP performance. Therefore, the ABEP of the SIC algorithm deceases with an increasing $X$.
\end{enumerate}

In addition, Fig. \ref{E2optimal} presents the $\Delta\epsilon_{\mathrm{opt}}$ against $X$, from which the following observations are made.
\begin{enumerate}
\item $\Delta\epsilon_{\mathrm{opt}}$ decreases with the increasing of $M$. When $M$ is large, the minimum Euclidean distance at the same latitude is small and the two circles of latitude have to be close enough to each other to offer an overall maximum $\Lambda_{\mathrm{V}}\Lambda_{\mathrm{H}}$.
\item $\Delta\epsilon_{\mathrm{opt}}$ increases with $X$. It is observed in Eq. (\ref{lambda1}) that if $X\approx0$, we have $\Lambda_{\mathrm{V}}\approx0$ when $k=\hat{k}$ and $q_{\mathrm{V}}=\hat{q}_{\mathrm{V}}$, so as to $\Lambda_{\mathrm{H}}$. Therefore, $\Delta \epsilon$ needs to be very small to guarantee a large enough $\Lambda_{\mathrm{V}}$. Hence, $\Delta\epsilon_{\mathrm{opt}}$ reduces with an decreasing $X$.
\item The impact of $X$ on $\Delta\epsilon_{\mathrm{opt}}$ decreases with an increasing $M$.
\end{enumerate}

\begin{figure} [!t]
\centering
\subfigure[6bpcu, C$_1$, $M=8$]{\includegraphics[width=1.5in]{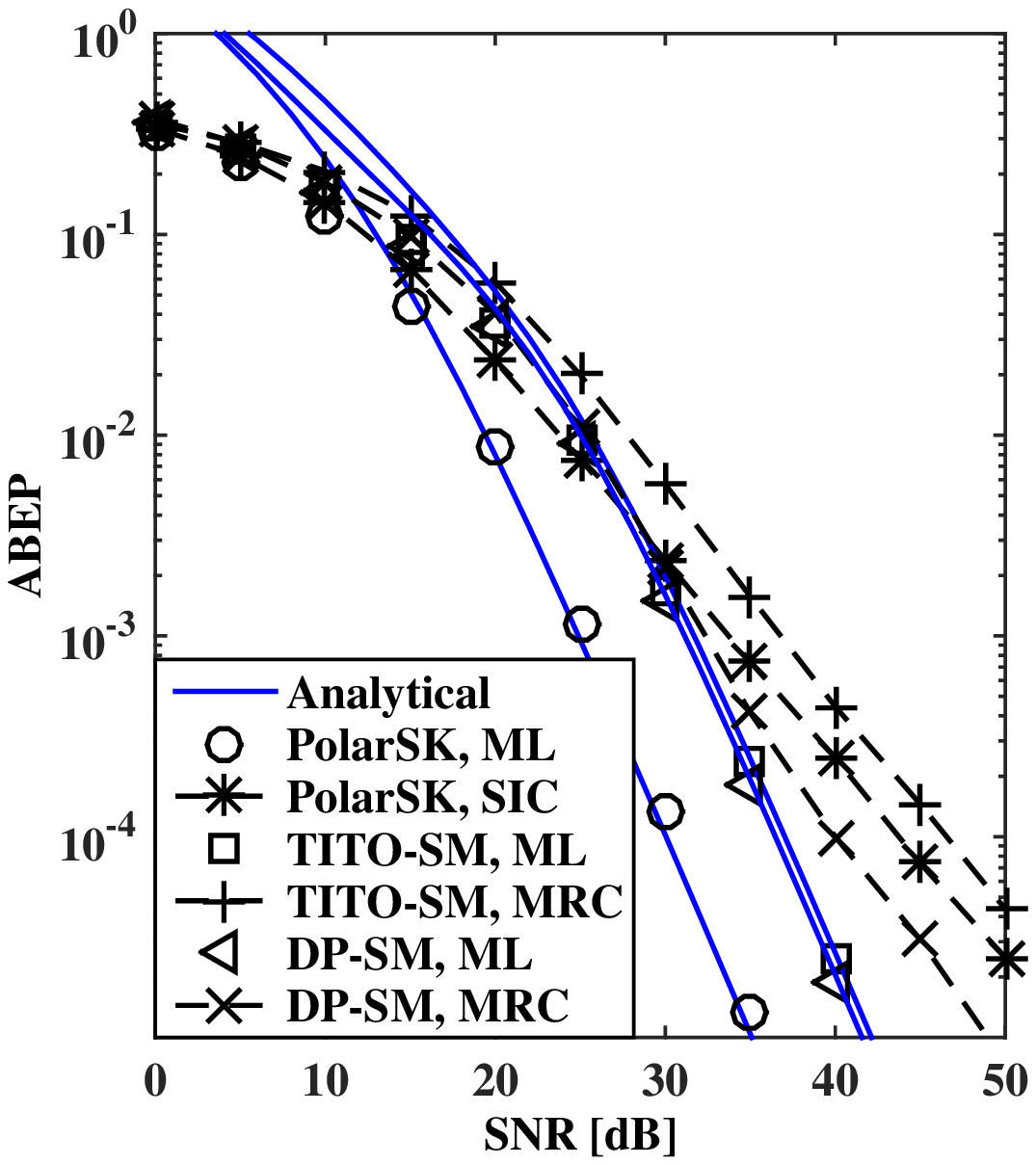}}
\subfigure[7bpcu, C$_2$, $M=8$]{\includegraphics[width=1.5in]{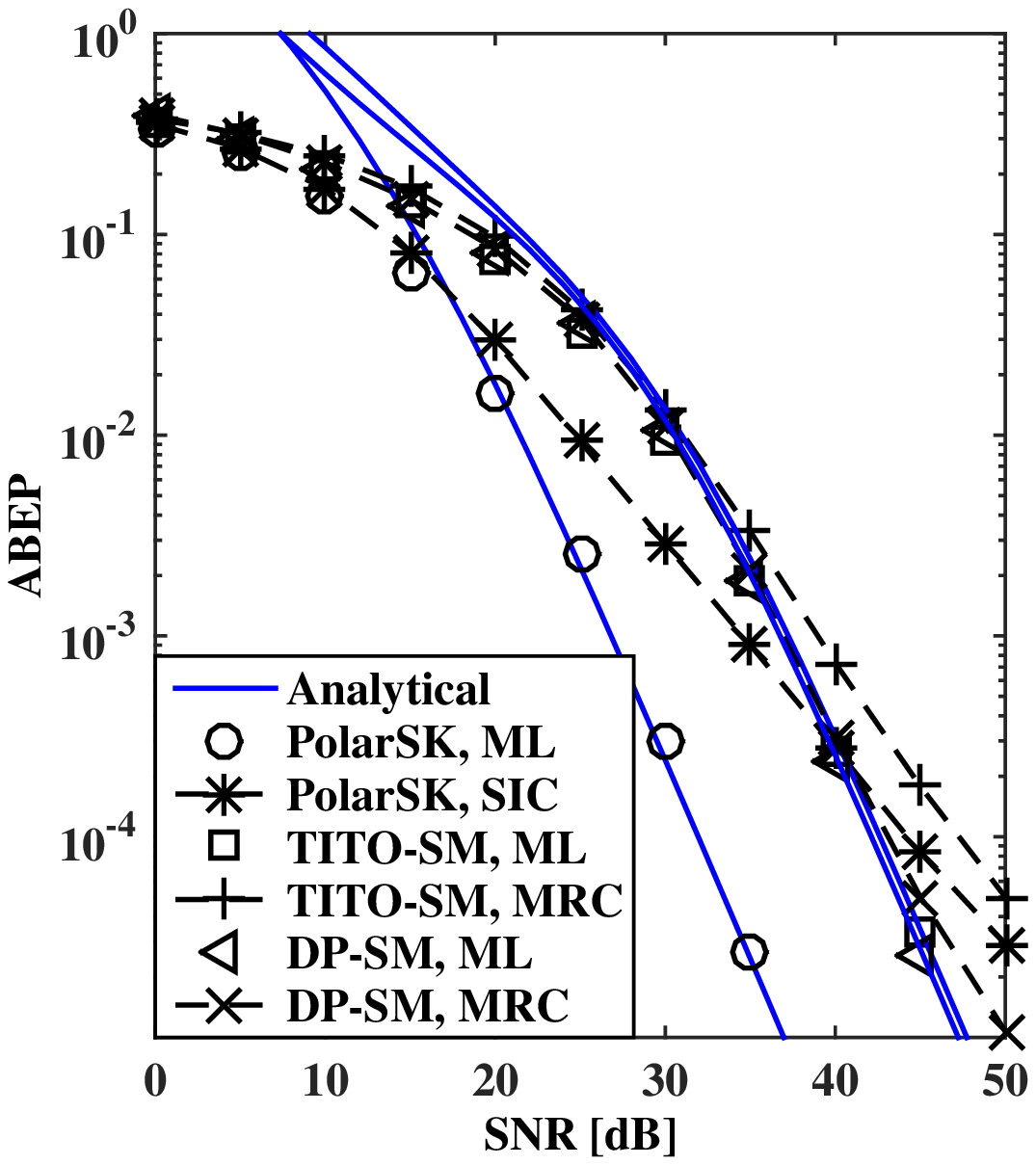}}
\caption{Comparison of PolarSK with DP-SM and TITO-SM systems for $N_{\mathrm{R}}=1$ under a same data rate. Markers show Monte Carlo simulations based on $2\times10^6$ realizations.
}\label{VSDPM}
  \vspace{-10pt}
\end{figure}

\subsection{Comparison to state of the art systems}

Recently, the DP-SM system and the UP TITO-SM system has been widely investigated.
In Fig. \ref{VSDPM}, a performance comparison of the PolarSK system against state of the art DP-SM and TITO-SM systems is provided.
For the DP-SM system and the TITO/TIMO system, the maximum ratio combining (MRC)-based linear detection algorithm \cite{MDRSSK3} is employed as a benchmark.
ABEP of the TITO-SM systems are computed by \cite{correlatesm}, and the Monte Carlo simulations are carried out for $2\times10^6$ channel realizations. For C$_2$ signal constellations, the optimal $\Delta \epsilon$ computed by (\ref{finaloptimal}) is used. For the TITO-SM system, the spatial correlation measured in \cite{mea3} in employed. $C_{\mathrm{T}}=\left[\begin{smallmatrix}1&0.83\\0.83&1\end{smallmatrix}\right],$
$C_{\mathrm{R}}=\left[\begin{smallmatrix}1&0.85\\0.85&1\end{smallmatrix}\right].$
The following observations can be made.

\begin{figure} [!t]
\centering
\subfigure[8bpcu. C$_1$, $M=16$]{\includegraphics[width=1.5in]{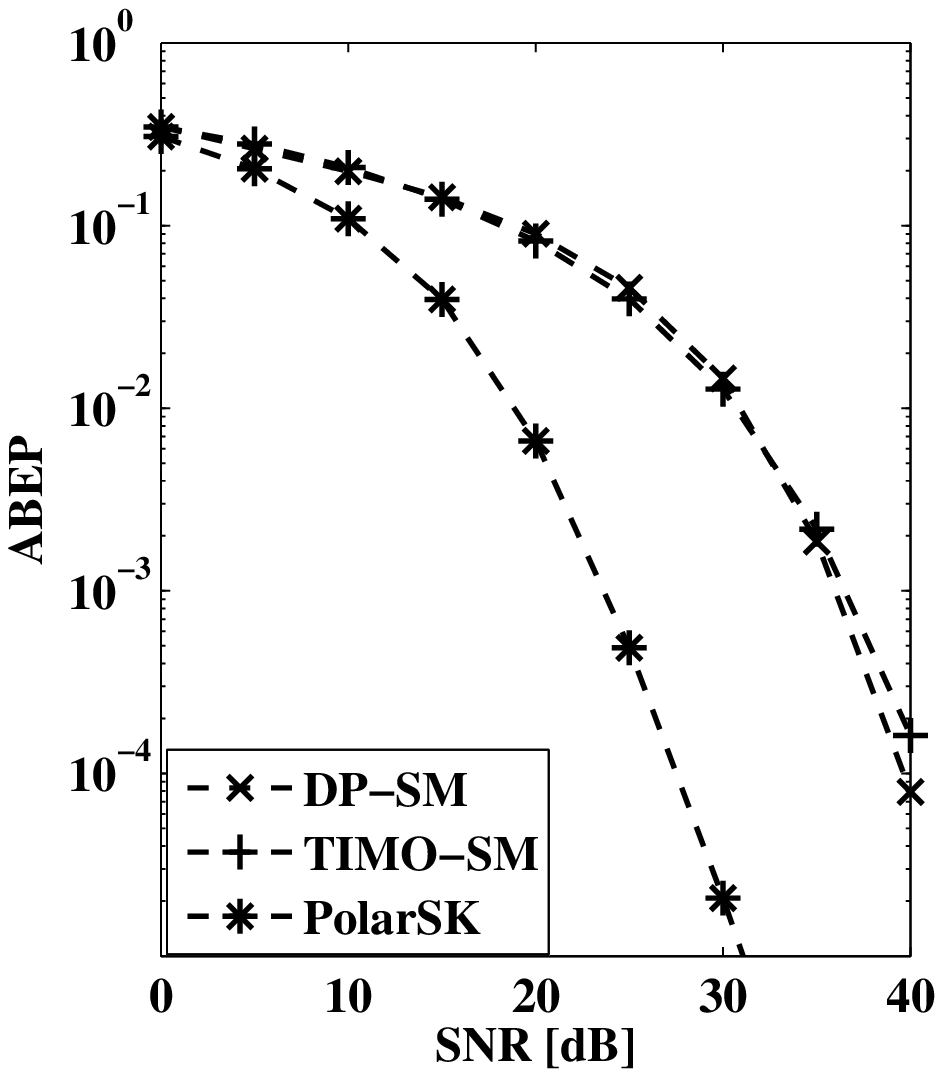}}
\subfigure[9bpcu. C$_2$, $M=16$]{\includegraphics[width=1.5in]{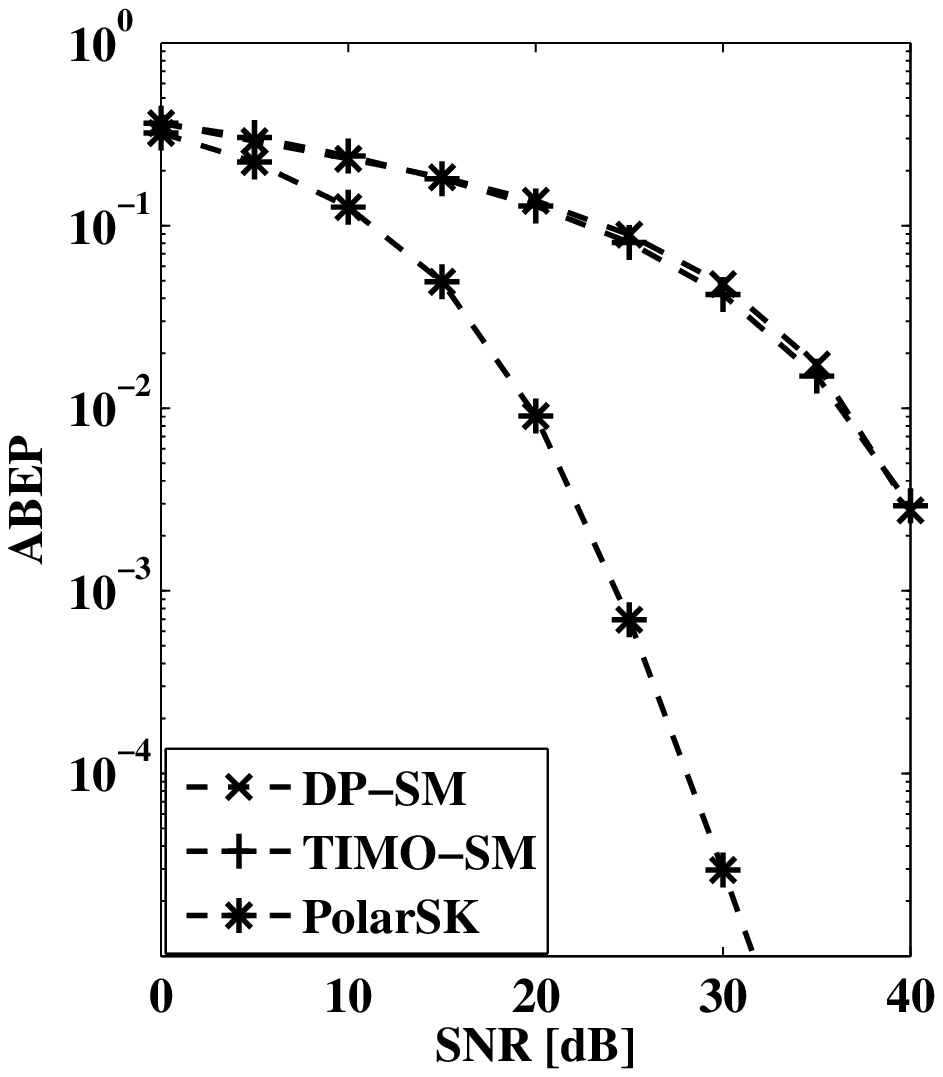}}
\caption{{Comparison of linear detection algorithms of PolarSK, DP-SM and TIMO-SM systems for $N_{\mathrm{R}}=2$ under a same data rate. ABEPs are obtained from Monte Carlo simulations based on $2\times10^6$ realizations.}
}\label{VSDPMhighorder}
  \vspace{-10pt}
\end{figure}

\begin{enumerate}
\item As predicted by (\ref{ABEPfinal2}), optimum detection algorithms of the PolarSK and the DP-SM systems achieve a same diversity order, i.e. $2N_{\mathrm{R}}$. As shown in \cite[Fig. 1-2]{correlatesm}, the diversity order of the convectional TITO/TIMO-SM system is $2N_{\mathrm{R}}$, too.
\item For the linear SIC receiver, because of superimposing independent information sequences to be transmitted by different polarized transmit antenna, the transmission of $q_{\mathrm{V}}$ and $q_{\mathrm{H}}$ interfere each other. Therefore, it is found in Fig. \ref{VSDPM} that the SIC algorithm of PolarSK system performs worse than the MRC DP-SM and the MRC TITO/TIMO-SM systems under high SNR regime. Nevertheless, the PolarSK system performs better than the DP-SM and the TITO-SM systems under low SNR regime.
\item For $N_{\mathrm{R}}=1$, we consider optimum receivers following Remark \ref{remcofig}. It is observed that both C$_1$ and C$_2$ PolarSK systems using SD detection algorithms outperform state of the art DP-SM and TITO-SM systems because the information that is mapped onto the polarization constellation can efficiently reduce the size of the signal constellation.
\end{enumerate}

Furthermore, ABEPs of the PolarSK, the DP-SM and the TIMO-SM against SNR are illustrated in Fig. \ref{VSDPMhighorder} for $N_{\mathrm{R}}=2$.
Following Remark \ref{remcofig}, only linear detection algorithms are taken into account. It is observed that the proposed PolarSK system using the linear SIC algorithm performs better than conventional DP-SM and TITO-SM systems significantly under practical values of SNR.

\subsection{Measurement analysis}
\label{measuresection}

Understanding the performance of the proposed PolarSK scheme from a theoretical perspective in a practical environment appears to be difficult due to the complexity of propagation channels. It is believed that only practical experiences can yield definitive answers on the achievable performance of single-RF system in real-world devices \cite{survey}. In this subsection, experimental results on the performance of PolarSK systems are for the first time given under a typical indoor scenario.

\begin{figure}[!t]
  \centering
  \includegraphics [width=2.3in]{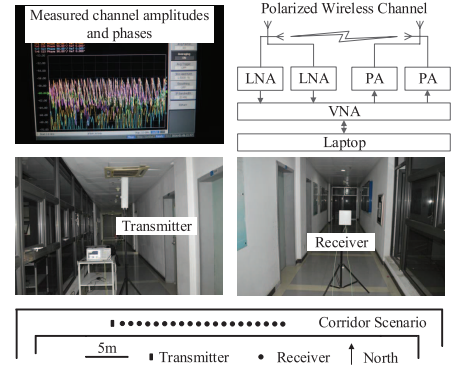}
  \centering\caption{Measurement system and scenario. {The diagram of the measurement system is in the upper right.}} \label{measurementscenario}
  \vspace{-10pt}
\end{figure}

In our work, a frequency domain channel sounder based on a {4-port} vector network analyzer (VNA), {which measures the S parameters of the wireless channel via discrete narrowband frequency tones swept across the bandwidth of interest,} is established. The transmitter and the receiver are both DP antennas that work at 2.4-2.5GHz. The gain of the transmit and the receive antennas are 12dBi and 14dBi, respectively. {Vertical and horizontal} receive antenna { elements are respectively} connected to { port 2 and port 1 of the VNA} by phase stable cables of 25 meters long, {whereas vertical and horizontal transmit antenna elements are respectively connected to port 3 and port 4 of the VNA.} The VNA is controlled by a laptop through the local area network (LAN). The major parts of the channel sounder and DP antennas are shown in  detail in Fig. \ref{measurementscenario}. {Measurement parameters are shown in TABLE \ref{measurementparameter}.  A description of the channel sounder along with a discussion of the conventional SM system can be found in \cite{mea3}.}

In a typical indoor scenario, as schematically depicted in Fig. \ref{measurementscenario}, both the transmitter and the receiver are located in a same building on the same floor. The measurement campaign is carried out on the second floor of the Building C of Harbin Institute of Technology Shenzhen Graduate School. The corridor has the dimension of 56m$\times$3.6m$\times$3m. There is one reference point of transmitter, noted by a rectangle marker, and 20 reference points of receivers, noted by circle markers.

We denote the measured $S$ parameters of the wireless channel as $\mathbf{H}_{0}(f, PL, r)$, where $f$ denotes the frequency, $PL\in\{\mathrm{VV},\mathrm{VH},\mathrm{HV},\mathrm{HH}\}$ denotes the polarization patterns of the transmit antenna and the receive antenna, and $r=1,2,...,20$ is the index of receiver reference point. More specifically, the  $\mathrm{S}_{31}$, $\mathrm{S}_{32}$, $\mathrm{S}_{41}$ and $\mathrm{S}_{42}$ parameters obtained by the VNA are respectively employed as
$\mathbf{H}_{0}(f,\mathrm{VH}, r)$,
$\mathbf{H}_{0}(f,\mathrm{VV}, r)$,
$\mathbf{H}_{0}(f,\mathrm{HH}, r)$, and $\mathbf{H}_{0}(f,\mathrm{HV}, r)$.

All measured channel transfer functions are normalized to their respective power as
\begin{eqnarray}
\setcounter{equation}{45}
\begin{array}{l}
\mathbf{H}(f, PL, r)=\frac{2B\mathbf{H}_0(f, PL, r)}{\sqrt{\int_f(|\mathbf{H}_0(f, \mathrm{VV}, r)|^2+|\mathbf{H}_0(f, \mathrm{HH}, r)|^2)\mathrm{d}f}}.
\end{array}
\end{eqnarray}
and $X$ under this scenario is estimated as
\begin{eqnarray}
\label{measuredx}
\begin{array}{l}
\hat{X}=\frac
{\sum\limits_r\int_f\{|\mathbf{H}(f, \mathrm{VH}, r)|^2+\mathbf{H}(f, \mathrm{HV}, r)|^2\}\mathrm{d}f}
{\sum\limits_r\int_f\{|\mathbf{H}(f, \mathrm{VV}, r)|^2+\mathbf{H}(f, \mathrm{HH}, r)|^2\}\mathrm{d}f}
\\
\hspace{0.18in}=-4.5665\mathrm{dB}.
\end{array}
\end{eqnarray}

\begin{figure}[!t]
\centering
\subfigure[4bpcu. C$_1$, $M=4$]{\includegraphics[width=1.5in]{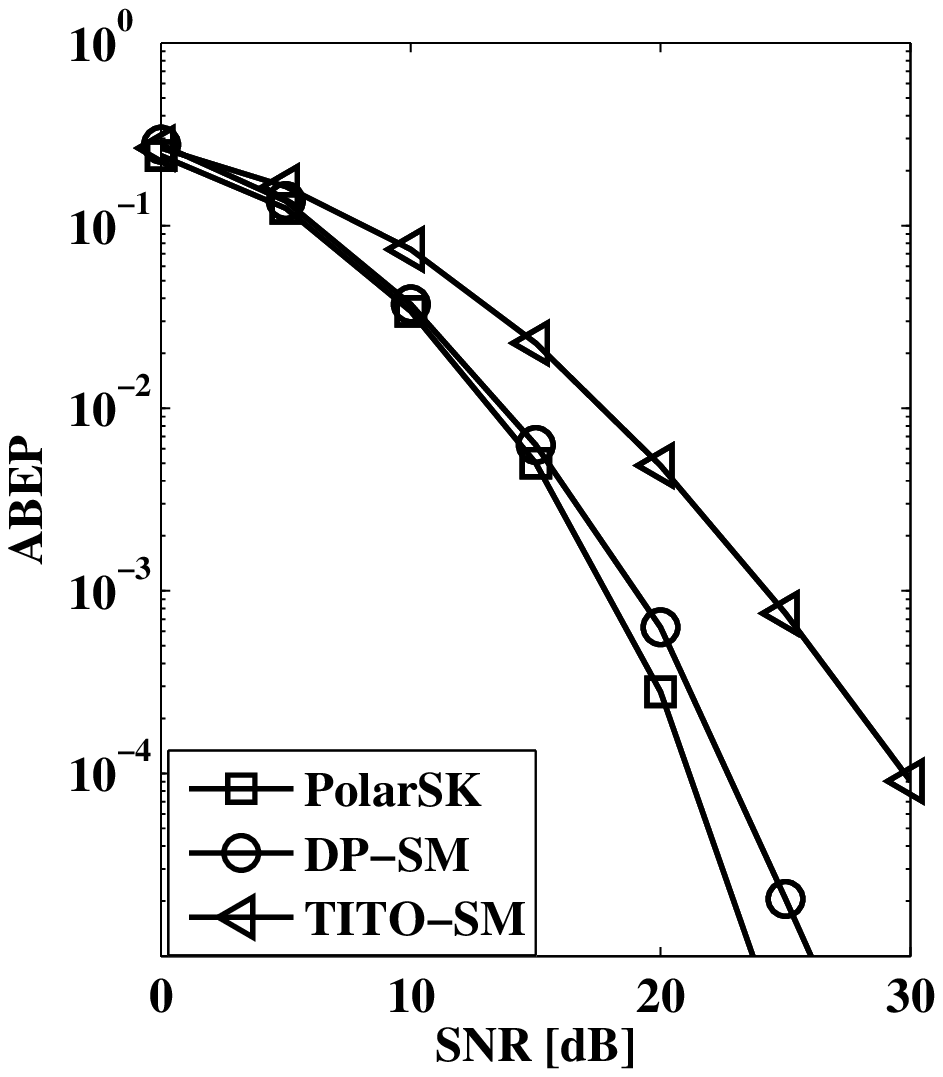}}
\subfigure[5bpcu. C$_2$. $M=4$]{\includegraphics[width=1.5in]{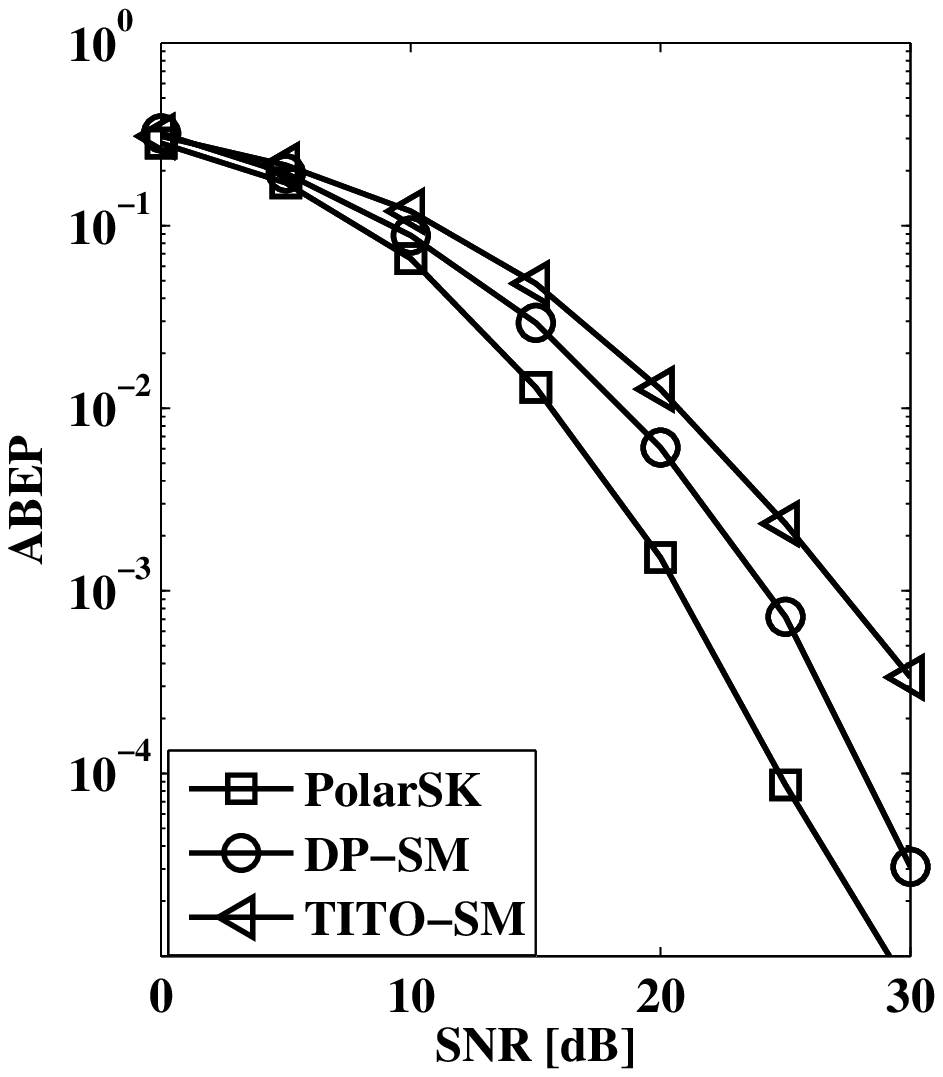}}
\caption{Comparison of the PolarSK, the DP-SM, and the TITO-SM systems for a same data rate under measured channels.
}\label{vssotamea}
  \vspace{-10pt}
\end{figure}

Over measured channels, we compare the ABEP of PolarSK with those of DP-SM and TITO-SM systems in Fig. \ref{vssotamea}.
The UP channel database in \cite{mea3} is employed to simulate the ABEP of the TITO-SM system. From Fig. \ref{vssotamea}, it is found that the proposed PolarSK system performs better than the DP-SM and the TITO-SM systems. 
Therefore, we conclude that the proposed PolarSK scheme outperforms the state of the art TITO-SM and DP-SM schemes in practical indoor environment.

\begin{table}[!t]
\centering
\caption{{Measurement parameters.}}
\label{measurementparameter}
\centering
\begin{tabular}{|l|l|}
\hline
Parameter&Value\\
\hline
Carrier frequency&2.45GHz\\
Band width&100MHz\\
Transmit power per polarization& 0dBm\\
Number of tones& 201\\
\hline
\end{tabular}
\end{table}

\section{Conclusions and future works}
\subsection{Conclusions}

In this paper, we have generalized the polarized single-RF MIMO scheme to a modulation technique referred to as the PolarSK scheme. The ML PolarSK receiver has been described and a closed-form ABEP upper bound under i.i.d. Rayleigh polarized channels has been derived. We have found that the diversity order of PolarSK system is $2N_{\mathrm{R}}$. On the basis of the analytic ABEP expression, the optimization of PolarSK signal constellation is presented. Numerical results show that the optimum signal constellation is able to minimize ABEPs of PolarSK systems. In order to facilitate the implementation of PolarSK systems, computationally efficient detection algorithms, i.e., the SIC receiver and the SD receiver are proposed. With a lower computational complexity than the optimum QR aided ML receiver, the SD receiver has the optimum ABEP performance. Whereas the SIC receiver has a higher ABEP and a lower computational complexity than the SD receiver. 
Numerical results have shown that the proposed PolarSK system outperforms DP-SM and UP TITO-SM systems in terms of ABEP.

\subsection{Future works}
The insights of this work suggest a further development of the following major research issues, which are grouped into two main topics: performance evaluation and system improvement.
Issues regarding performance are as follows.
\begin{figure*}[bp]
  \vspace{-10pt}
\hrulefill
\begin{eqnarray}
\setcounter{equation}{52}
\label{pepdef2}
\begin{array}{l}
\mathrm{Pr}(p\rightarrow \hat{p}) =\mathrm{E}\left[Q\left(\sqrt{\frac{\rho}{2}
\left[
\begin{smallmatrix}
\sum\limits_{n_{\mathrm{R}}=1}^{N_{\mathrm{R}}}
|\Delta x_{\mathrm{V}} h_{\mathrm{VV},n_{\mathrm{R}}} +\sqrt{X}\Delta x_{\mathrm{H}}h_{\mathrm{VH},n_{\mathrm{R}}}|^2\\
+
\sum\limits_{n_{\mathrm{R}}=1}^{N_{\mathrm{R}}}
|\Delta x_{\mathrm{H}} h_{\mathrm{HH},n_{\mathrm{R}}} +\sqrt{X}\Delta x_{\mathrm{V}}h_{\mathrm{HV},n_{\mathrm{R}}}|^2
\end{smallmatrix}
\right]
}\right)\right]
\\ \hspace{0.67in}
=\mathrm{E}\left[Q\left(\sqrt{\frac{\rho}{4}
\left[
\underbrace{(|\Delta x_{\mathrm{V}}|^2+X|\Delta x_{\mathrm{H}}|^2)}_{\Lambda_{\mathrm{V}}}\chi_{2N_{\mathrm{R}},1}^2
+
\underbrace{(|\Delta x_{\mathrm{H}}|^2+X|\Delta x_{\mathrm{V}}|^2)}_{\Lambda_{\mathrm{H}}}\chi_{2N_{\mathrm{R}},2}^2
\right]
}\right)\right].
\end{array}
\end{eqnarray}
\end{figure*}

\begin{enumerate}
   \item To deeper understand PolarSK systems, analytical analysis  over generalized fading channels, such as spatial correlated Ricean channel, Nakagami-m channel {\cite{bepsm}} and keyhole channel {\cite{keyholeSM}}, is required. Also, to investigate how PolarSK performs under more realistic channels, it is necessary to evaluate system performances under statistic channel models such as SCM  {\cite{3GPP}}.
   \item In this paper, we have assumed that the receiver has full channel state information. In the future, the robustness of PolarSK systems to channel state information errors needs to be evaluated.
   \item In this future, the performance of PolarSK systems will be measured under more typical scenarios. On the basis of analytic evaluation scheme of PolarSK under generalized channel models, a systematic comparison of analytic and measured results is crucial to  yield definitive answers to the performance of PolarSK systems.
\end{enumerate}

Issues regarding system improvement are as follows.
\begin{enumerate}
   \item Using tri-polarized antenna where the composite structure is composed of three orthogonal sleeves, an extra factor of three in channel capacity can be obtained, relative to the conventional limit of using single polarized antenna  {\cite{tripolar}}. The capacity of PolarSK will be further enhanced.
   \item An open research issue for PolarSK is the design of encoding and decoding schemes that exploit the spatial and polarization constellation diagrams to achieve transmit diversity gains. Previously, the SM scheme which can achieve transmit diversity gains was well designed \cite{im1}. When using polarization resources, allocation algorithms for balancing diversity gains and multiplexing gains in both space domain and polarization domain shall be further investigated.
    \item It is of interest to generalize the PolarSK system for a generic $N_{\mathrm{T}} \times N_{\mathrm{R}}$ MIMO scheme, where $N_{\mathrm{T}}$ and $N_{\mathrm{R}}$ denote the numbers of dual-polarized antennas at the transmitter and receiver, respectively.
   \item In this paper, the PolarSK scheme is analyzed in the link level. In the future, the PolarSK scheme needs be applied to various networks, e.g., HetNets and relay networks. Especially in the HetNets, $X$ is heterogeneous under various environments, which bring challenge to the optimization of signal constellation.
\end{enumerate}

\ifCLASSOPTIONcaptionsoff
  \newpage
\fi
\appendices

\section{Proof of Lemma \ref{peplemma}}
\label{prooflemma0appen}

For each pair of $\mathbf{x}_p$ and $\mathbf{x}_{\hat{p}}$, {following the definition of PEP and according to Eq. (\ref{mlreceiver})}, we have
\begin{eqnarray}
\setcounter{equation}{47}
\label{pep0}
\begin{array}{l}
\mathrm{Pr}(p\rightarrow \hat{p})=\mathrm{Pr}(\|\mathbf{y}-\mathbf{H}\mathbf{x}_p\|^2>\|\mathbf{y}-\mathbf{H}\mathbf{x}_{\hat{p}}\|^2).
\end{array}
\end{eqnarray}
{
Substituting (\ref{receivedsignal}) into (\ref{pep0}), we have
\begin{eqnarray}
\begin{array}{l}
\mathrm{Pr}(p\rightarrow \hat{p})=\mathrm{Pr}(\|\frac{\mathbf{w}}{\sqrt{\rho}}\|^2>\|\mathbf{H}(\mathbf{x}_{p}-\mathbf{x}_{\hat{p}})+\frac{\mathbf{w}}{\sqrt{\rho}}\|^2)
\\\hspace{0.66in}=\mathrm{Pr}\left(\begin{smallmatrix}-2\mathrm{Re}[\mathbf{w}^H\mathbf{H}(\mathbf{x}_{p}-\mathbf{x}_{\hat{p}})]
\\
>\sqrt{\rho}\|\mathbf{H}(\mathbf{x}_{p}-\mathbf{x}_{\hat{p}})\|^2
\end{smallmatrix}
\right).
\end{array}
\end{eqnarray}
Since $\mathbf{w}^H\mathbf{H}(\mathbf{x}_{p}-\mathbf{x}_{\hat{p}})\sim \mathcal{CN}\left(0,{\|\mathbf{H}(\mathbf{x}_{p}-\mathbf{x}_{\hat{p}})\|^2}\right)$, we have $-2\mathrm{Re}[\mathbf{w}^H\mathbf{H}(\mathbf{x}_{p}-\mathbf{x}_{\hat{p}})]\sim \mathcal{N}\left(0,2\|\mathbf{H}(\mathbf{x}_{p}-\mathbf{x}_{\hat{p}})\|^2\right)$
 and therefore
\begin{eqnarray}
\begin{array}{l}
\mathrm{Pr}(p\rightarrow \hat{p})=\mathrm{Pr}\left(\begin{smallmatrix}\mathcal{N}\left(0,2\|\mathbf{H}(\mathbf{x}_{p}-\mathbf{x}_{\hat{p}})\|^2\right)
\\
>\sqrt{\rho}\|\mathbf{H}(\mathbf{x}_{p}-\mathbf{x}_{\hat{p}})\|^2
\end{smallmatrix}
\right)
\\
\hspace{0.66in}=\mathrm{Pr}\left(\mathcal{N}\left(0,1\right)
>\sqrt{\frac{\rho}{2}\|\mathbf{H}(\mathbf{x}_{p}-\mathbf{x}_{\hat{p}})\|^2}
\right).
\end{array}
\end{eqnarray}
}
{Following the definition of the Q-function,} we obtain
\begin{eqnarray}
\label{pepdef}
\mathrm{Pr}(p\rightarrow \hat{p})=\mathrm{E}\left[Q\left(\sqrt{\frac{\rho}{2}\left\|\mathbf{H}\Delta\mathbf{x}(p,\hat{p})\right\|^2}\right)\right],
\end{eqnarray}
where
\begin{eqnarray}
\label{costel}
\begin{array}{l}
\Delta\mathbf{x}(p,\hat{p})
=
\left[
\begin{smallmatrix}
\Delta x_{\mathrm{V}}(k,q_{\mathrm{V}},\hat{k},\hat{q}_{\mathrm{V}})\\
\Delta x_{\mathrm{H}}(k,q_{\mathrm{H}},\hat{k},\hat{q}_{\mathrm{H}})
\end{smallmatrix}
\right]\\
\hspace{0.54in}\triangleq
\left[
\begin{smallmatrix}
\cos\epsilon_{k}\exp\left(j\frac{2\pi q_{\mathrm{V}}}{ M}\right)-\cos\epsilon_{\hat{k}}\exp\left(j\frac{2\pi\hat{q}_{\mathrm{V}}}{ M}\right)
\\
\sin\epsilon_{k}\exp\left(j\frac{2\pi q_{\mathrm{H}}}{ M}\right)-\sin\epsilon_{\hat{k}}\exp\left(j\frac{2\pi \hat{q}_{\mathrm{H}}}{ M}\right)
\end{smallmatrix}
\right].
\end{array}
\end{eqnarray}

{

Substituting (\ref{channelmodeleq})-(\ref{channelmodeleq2}) into (\ref{pepdef}), we have (\ref{pepdef2}).
Substituting $Q(x)\equiv \frac{1}{\pi} \int_0^{\frac{\pi}{2}} \exp \left( - \frac{x^2}{2 \sin^2 \theta} \right) \mathrm{d}\theta$ into (\ref{pepdef2}), we have
\begin{eqnarray}
\setcounter{equation}{53}
\label{pepdef3}
\begin{array}{l}
\mathrm{Pr}(p\rightarrow \hat{p})
=\frac{1}{\pi} \int_0^{\frac{\pi}{2}}{
\mathcal{M}_{\Lambda_{\mathrm{V}}\chi_{2N_{\mathrm{R}},1}^2
+\Lambda_{\mathrm{H}}\chi_{2N_{\mathrm{R}},2}^2}\left(\frac{\rho}{8\sin^2 \theta}\right)\mathrm{d}\theta}
\\
=\frac{1}{\pi} \int_0^{\frac{\pi}{2}}{
\mathcal{M}_{\Lambda_{\mathrm{V}}\chi_{2N_{\mathrm{R}},1}^2}\left(\frac{\rho}{8\sin^2 \theta}\right)
\mathcal{M}_{\Lambda_{\mathrm{H}}\chi_{2N_{\mathrm{R}},2}^2}\left(\frac{\rho}{8\sin^2 \theta}\right)
\mathrm{d}\theta}
\end{array}
\end{eqnarray}
where $\mathcal{M}_{Z}(s)\triangleq \mathrm{E}\left[\exp(-sZ)\right]$.
Since
\begin{eqnarray}
&&\mathcal{M}_{\Lambda_{\mathrm{V}}\chi_{2N_{\mathrm{R}},1}^2}\left(s\right)
=(1+2\Lambda_{\mathrm{V}}s)^{-N_{\mathrm{R}}},
\\
&&\mathcal{M}_{\Lambda_{\mathrm{H}}\chi_{2N_{\mathrm{R}},2}^2}\left(s\right)
=(1+2\Lambda_{\mathrm{H}}s)^{-N_{\mathrm{R}}},
\end{eqnarray}
we obtain,
\begin{eqnarray}
\label{pepeq1}
\begin{array}{l}
\mathrm{Pr}(p\rightarrow \hat{p})
=\frac{1}{\pi}\int_{0}^{\frac{\pi}{2}}
\left[
\frac{\sin^4\theta}{(\sin^2\theta+\frac{\rho\Lambda_{\mathrm{V}}}{4})(\sin^2\theta+\frac{\rho\Lambda_{\mathrm{H}}}{4})}
\right]^{N_{\mathrm{R}}}
\mathrm{d}\theta.
\end{array}
\end{eqnarray}
Changing the variable $t=\cos^2\theta$ in (\ref{pepeq1}), with some mathematical manipulations, we derive a closed form of the exact PEP as (\ref{peprayeq5}).
}

\section{Proof of Lemma \ref{d1}}
\label{prooflemma1appen}
{ As shown in Eq. (\ref{optcondition}), the goal of the optimization problem is to maximize $\min\{\Lambda_{\mathrm{V}}\Lambda_{\mathrm{H}}\}$. Clearly, it is straightforwardly observed from both extreme conditions in Section V that  $\min\{\Lambda_{\mathrm{V}}\Lambda_{\mathrm{H}}\}$ decreases with an increasing $\Delta\epsilon$ for $\hat{k}=k$, whereas $\min\{\Lambda_{\mathrm{V}}\Lambda_{\mathrm{H}}\}$ increases with $\Delta\epsilon$ for $\hat{k}\neq k$. To maximize the overall $\min\{\Lambda_{\mathrm{V}}\Lambda_{\mathrm{H}}\}$, we have to derive it respectively under conditions of $\hat{k}=k$ and $\hat{k}\neq k$ firstly.}

If $\hat{k}=k$, by substituting (\ref{costel}) into Eqs. (\ref{lambda1})-(\ref{lambda2}), we have
\begin{eqnarray}
\label{lambda1e2}
&&
\hspace{-40pt}
\begin{array}{l}
\Lambda_{\mathrm{V}}=\cos^2({\epsilon_{\mathrm{max}}})\left|\exp\left(j\frac{2\pi q_{\mathrm{V}}}{M}\right)-\exp\left(j\frac{2\pi \hat{q}_{\mathrm{V}}}{M}\right)\right|^2\\ \hspace{17pt}
+X\sin^2({\epsilon_{\mathrm{max}}})\left|\exp\left(j\frac{2\pi q_{\mathrm{H}}}{M}\right)-\exp\left(j\frac{2\pi \hat{q}_{\mathrm{H}}}{M}\right)\right|^2,
\end{array}\\
\label{lambda2e2}
&&
\hspace{-40pt}
\begin{array}{l}
\Lambda_{\mathrm{H}}=X\cos^2({\epsilon_{\mathrm{max}}})\left|\exp\left(j\frac{2\pi q_{\mathrm{V}}}{M}\right)-\exp\left(j\frac{2\pi \hat{q}_{\mathrm{V}}}{M}\right)\right|^2\\ \hspace{17pt}
+\sin^2({\epsilon_{\mathrm{max}}})\left|\exp\left(j\frac{2\pi q_{\mathrm{H}}}{M}\right)-\exp\left(j\frac{2\pi \hat{q}_{\mathrm{H}}}{M}\right)\right|^2.
\end{array}
\end{eqnarray}

For any $q\neq \hat{q}$, we have $\left|\exp\left(j\frac{2\pi q}{M}\right)-\exp\left(j\frac{2\pi \hat{q}}{M}\right)\right|^2\geq \left|1-\exp\left(j\frac{2\pi}{M}\right)\right|^2=2-2\cos\left(\frac{2\pi}{M}\right)$. Because $(k,q_{\mathrm{V}},q_{\mathrm{H}})\neq(\hat{k},\hat{q}_{\mathrm{V}},\hat{q}_{\mathrm{H}})$, we have $(q_{\mathrm{V}},q_{\mathrm{H}})\neq(\hat{q}_{\mathrm{V}},\hat{q}_{\mathrm{H}})$ when $\hat{k}=k$. Without loss of generality, under the condition $\hat{q}_{\mathrm{H}}\neq {q}_{\mathrm{H}}$, we obtain
\begin{eqnarray}
\Lambda_{\mathrm{V}}\Lambda_{\mathrm{H}}\geq {\cos^4}\left({\epsilon_{\mathrm{max}}}\right)\left[2-2\cos\left(\frac{2\pi}{M}\right)\right]^2,
\end{eqnarray}
where $\epsilon_{\mathrm{max}}\triangleq \max\limits_{k}\{\epsilon_{k},\frac{\pi}{2}-\epsilon_{k}\}$.
Some algebraic manipulations leads to
\begin{eqnarray}
\label{minilambv1}
\begin{array}{l}
\min\{\Lambda_{\mathrm{V}}\Lambda_{\mathrm{H}}|k=\hat{k}\}= X(1{+\cos2\epsilon_{\mathrm{max}}})^2
\\
\hspace{1.15in}\times\left(1-\cos\frac{2\pi}{M}\right)^2.
\end{array}
\end{eqnarray}

Else if $\hat{k}\neq k$, by using inequalities
\begin{eqnarray}
\label{ineq1}
&&\hspace{-20pt}
\begin{array}{l}
\left|\begin{smallmatrix}\cos(\epsilon_{k_1})\exp\left(j\frac{2\pi q_{\mathrm{V}}}{M}\right)\\-\cos(\epsilon_{k_2})\exp\left(j\frac{2\pi \hat{q}_{\mathrm{V}}}{M}\right)\end{smallmatrix}\right|
\geq\left|\cos(\epsilon_{k_1})-\cos(\epsilon_{k_2})\right|,
\end{array}
\\
\label{ineq2}
&&\hspace{-20pt}
\begin{array}{l}
\left|\begin{smallmatrix}\sin(\epsilon_{k_1})\exp\left(j\frac{2\pi q_{\mathrm{H}}}{M}\right)
\\
-\sin(\epsilon_{k_2})\exp\left(j\frac{2\pi \hat{q}_{\mathrm{H}}}{M}\right)\end{smallmatrix}\right|
\geq\left|\sin(\epsilon_{k_1})-\sin(\epsilon_{k_2})\right|,
\end{array}
\end{eqnarray}
we have
{
\begin{eqnarray}
\label{lambda1e22}
&&\hspace{-40pt}\begin{array}{l}
\Lambda_{\mathrm{V}}\geq
\left|\cos(\epsilon_{k_1})-\cos(\epsilon_{k_2})\right|^2+X\left|\sin(\epsilon_{k_1})-\sin(\epsilon_{k_2})\right|^2,
\end{array}\\
\label{lambda2e22}
&&\hspace{-40pt}\begin{array}{l}
\Lambda_{\mathrm{H}}\geq X\left|\cos(\epsilon_{k_1})-\cos(\epsilon_{k_2})\right|^2+\left|\sin(\epsilon_{k_1})-\sin(\epsilon_{k_2})\right|^2.
\end{array}
\end{eqnarray}}
Some algebraic manipulations leads to
\begin{eqnarray}
\label{minilambv2}
\begin{array}{l}
\min\{\Lambda_{\mathrm{V}}\Lambda_{\mathrm{H}}|k\neq\hat{k}\}=
(1+X)^2\sin^4\Delta\epsilon.
\end{array}
\end{eqnarray}

Combining Eqs. (\ref{minilambv1}) and (\ref{minilambv2}), we have the minimum $\Lambda_{\mathrm{V}}\Lambda_{\mathrm{H}}$ expression for arbitrary $M$ and $\epsilon_k$ as
\begin{eqnarray}
\min\left\{\min\{\Lambda_{\mathrm{V}}\Lambda_{\mathrm{H}}|k=\hat{k}\},\min\{\Lambda_{\mathrm{V}}\Lambda_{\mathrm{H}}|k\neq\hat{k}\}\right\}.
\end{eqnarray}

Accordingly, the problem is equivalent to selecting the $\Delta\epsilon_{\mathrm{opt}}$ to let
\begin{eqnarray}
\label{eq29}
\begin{array}{l}
X\left(1{+\cos2\epsilon_{\mathrm{max}}}\right)^2\left(1-\cos\frac{2\pi}{M}\right)^2=\left(1+X\right)^2\sin^4\Delta\epsilon.
\end{array}
\end{eqnarray}

For $K>1$, since $\epsilon_{\mathrm{max}}\geq \frac{\pi}{4}+ (K-1)\Delta\epsilon$, and with a given $\Delta\epsilon$, the system performance increases with a reducing $\epsilon_{\mathrm{max}}$, we use
signal constellation with $\epsilon_{\mathrm{max}}= \frac{\pi}{4}+(K-1)\Delta\epsilon$. Substituting $\epsilon_{\mathrm{max}}= \frac{\pi}{4}+(K-1)\Delta\epsilon$ into (\ref{eq29}) and following some algebraic manipulations, we obtain (\ref{finaloptimal}).

\begin{IEEEbiography}
[{\includegraphics[width=1in,height=1.25in,clip,keepaspectratio]{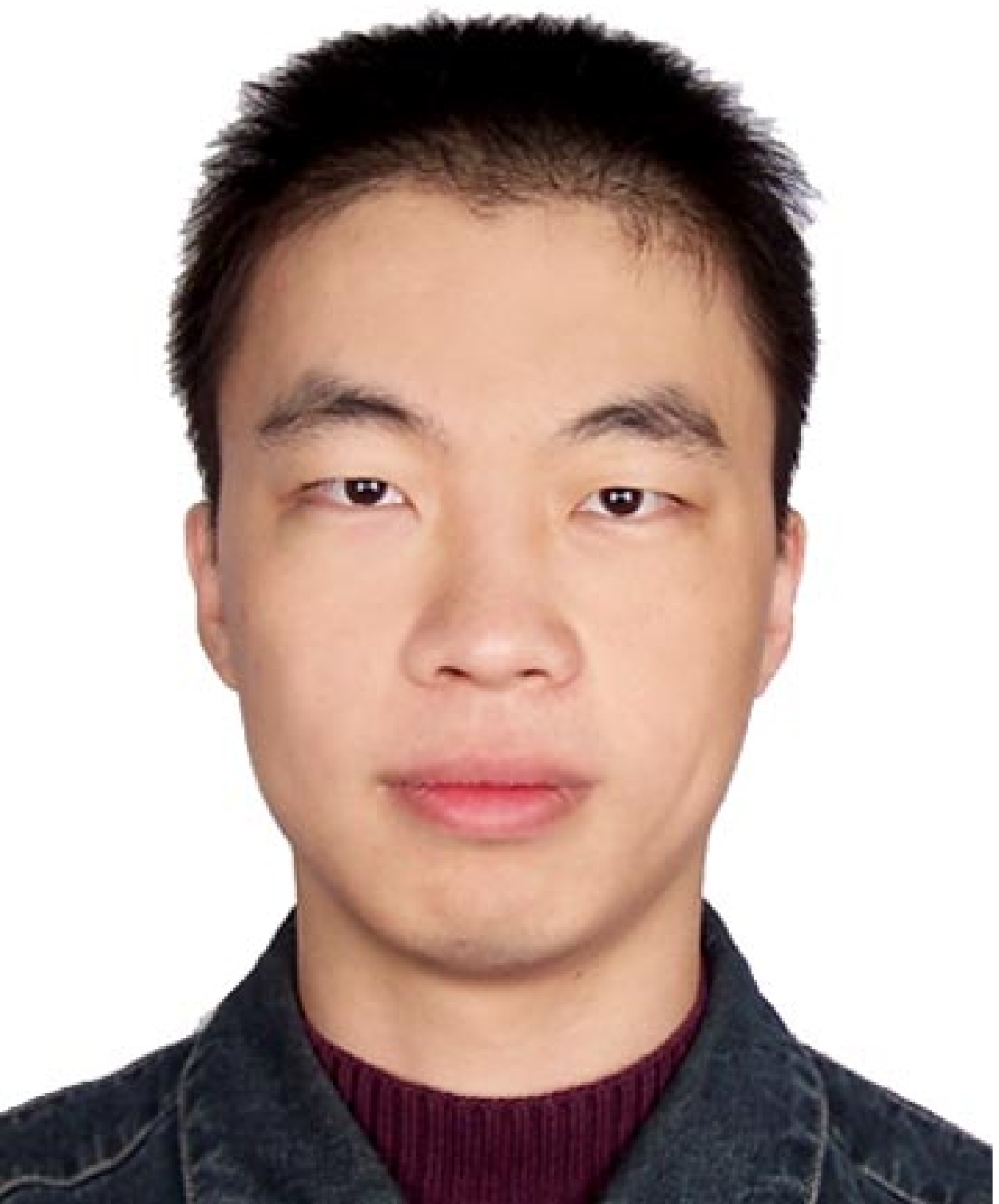}}]
{Jiliang Zhang} (\textbf{M}'15) is currently an associate professor at School of Information Science and Engineering, Lanzhou University. He received the B.S., M.S. and Ph.D. degrees from the Harbin Institute of Technology in 2007, 2009 and 2014, respectively. His research interests cover a wide range of topics in wireless systems, in particular including MIMO channel measurement and modelling, single radio frequency MIMO system, lattice coding design, full-duplex relay system, and wireless ranging system.
\end{IEEEbiography}

\begin{IEEEbiography}
[{\includegraphics[width=1in,height=1.25in,clip,keepaspectratio]{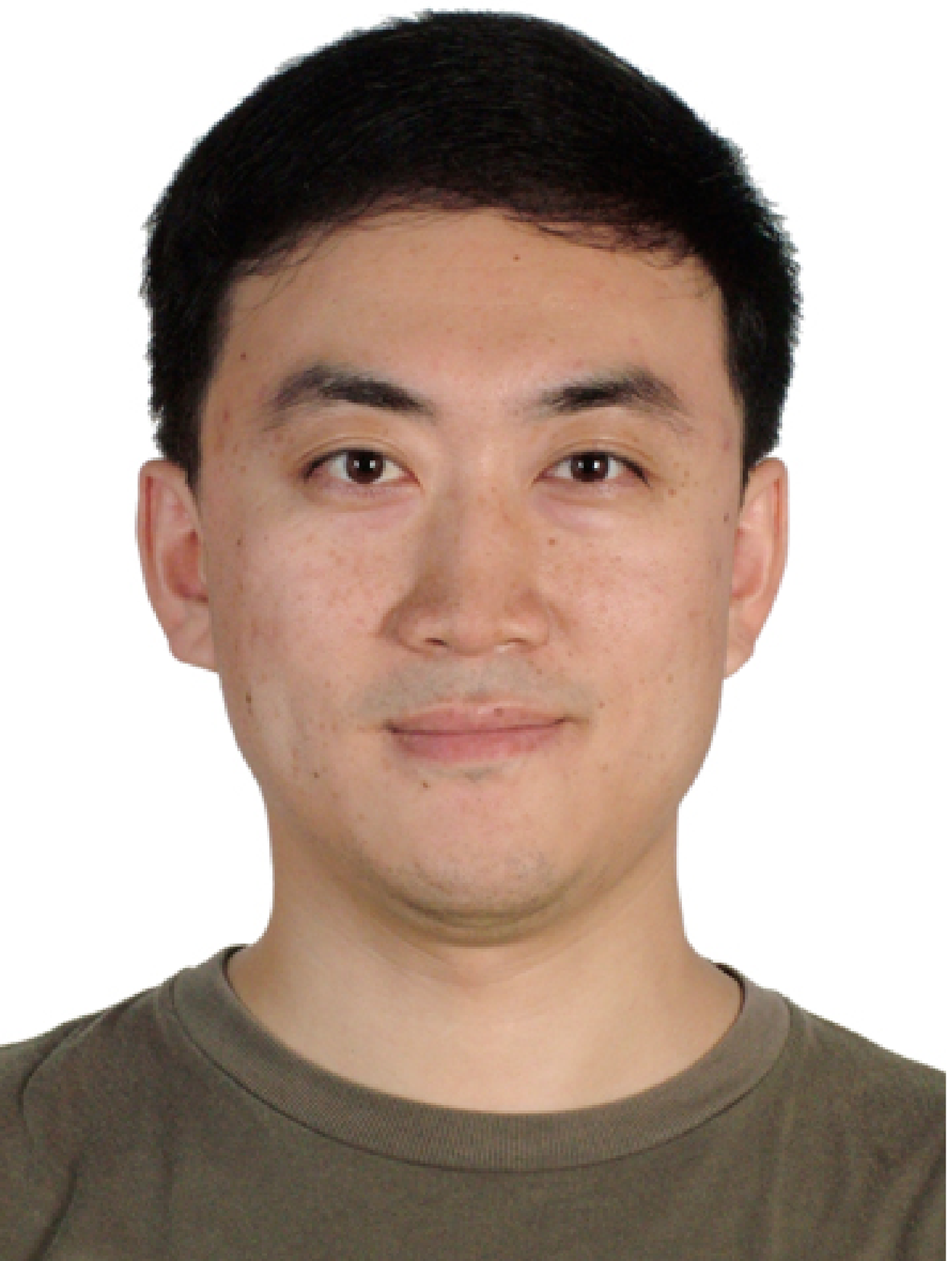}}]
{Yang Wang}  received his Ph.D. degree from Harbin Institute of Technology in 2005. From 2005 to 2007, he was a postdoctoral at Harbin Institute of Technology Shenzhen Graduate School, Shenzhen, China. He is currently an associate professor and doctoral supervisor of Harbin Institute of Technology Shenzhen Graduate School. He is a senior member of Chinese Institute of Electronics. His research interests include wireless communications, UWB system, MIMO system, signal processing and cooperative communications.
\end{IEEEbiography}

\begin{IEEEbiography}
[{\includegraphics[width=1in,height=1.25in,clip,keepaspectratio]{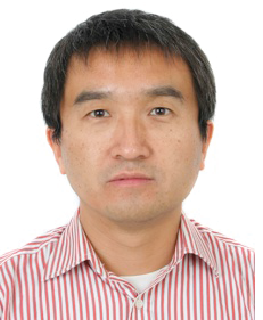}}]
{Jie Zhang}(\textbf{M}'05-\textbf{SM}'16)  received the Ph.D. degree in automatic control and electronic engineering from East China University of Science and Technology, Shanghai, China, in 1995.

He has been a Full Professor and the Chair in Wireless Systems with the Department of Electronic and Electrical Engineering, University of Sheffield, Sheffield, U.K., since January 2011. From January 2002 to January 2011, he was with the University of Bedfordshire, where he became a Lecturer, Reader, and Professor in 2002, 2005, and 2006, respectively. With his students/colleagues, he has pioneered research in femto/small cell and HetNets and published some of the earliest and/or most cited publications in these topics. Since 2005, he has been awarded more than 20 grants by the Engineering and Physical Sciences Research Council, the EC FP6/FP7/H2020, and industry, including some of world's earliest research projects on femtocell/HetNets. He cofounded Ranplan Wireless Network Design Ltd., Cambridge, U.K., which produces a suite of world-leading in-building distributed antenna systems, indoor-outdoor small cell/HetNet network planning, and optimization tools \textit{iBuildNet}$^\circledR$ that have been used by Ericsson, Huawei, and Cisco.
\end{IEEEbiography}

\begin{IEEEbiography}
[{\includegraphics[width=1in,height=1.25in,clip,keepaspectratio]{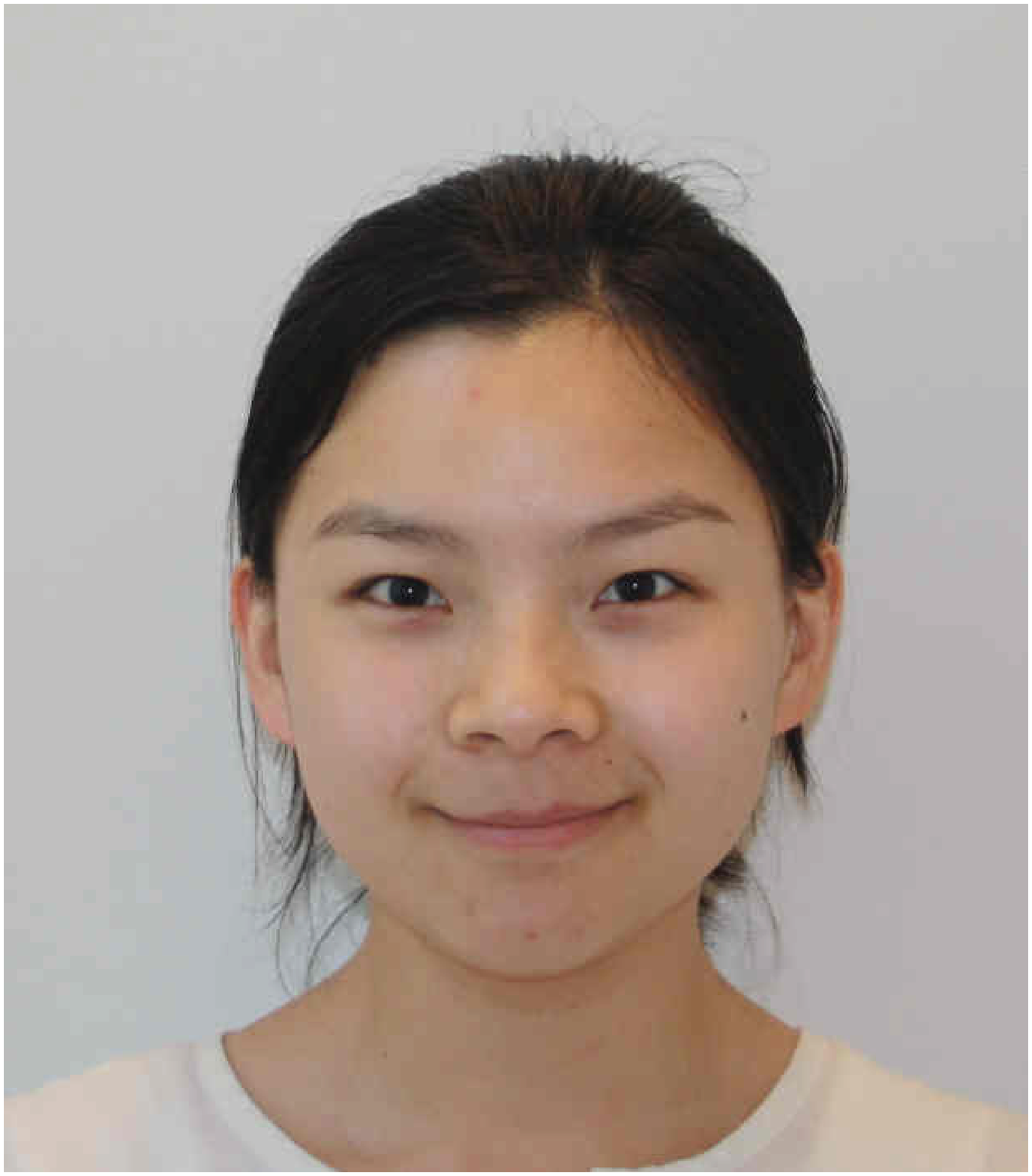}}]
{Liqin Ding} (\textbf{M}'17) received the B.S., M.S., and Ph.D degree in Information and Communication Engineering from Harbin Institute of Technology, Harbin, China, in 2009, 2011, and 2017, respectively. She was visiting the Department of Electronics and Telecommunications, Norwegian University of Science and
Technology from Sept. 2012 to Sept. 2014. She is currently a postdoctoral fellow in Information and Communication Engineering at Shenzhen Graduate School, Harbin Institute of Technology, Shenzhen, China. Her research interests include lattice theory, signal processing and cooperative design for wireless communications.
\end{IEEEbiography}

\end{document}